\documentclass{jfm}
\usepackage{graphicx}
\usepackage{epstopdf, epsfig}
\usepackage{amsmath}
\usepackage{xcolor}
\usepackage{framed}

\newcommand{\LSF}{{\mathrm L}}
\newcommand{\DELT}{\Delta\!}
\newcommand{\pd}[2]{\frac{\partial #1}{\partial #2}}

\shorttitle{Extension of QSQH theory}
\shortauthor{S. Chernyshenko}

\title{Extension of QSQH theory of scale interaction in near-wall turbulence to all velocity components}

\author{Sergei Chernyshenko
  \corresp{\email{s.chernyshenko@imperial.ac.uk}}}

\affiliation{Department of Aeronautics, Imperial College London, London SW7 2AZ, UK}

\begin{document}

\maketitle

\newcommand{\revision}[1]{{#1}}
\newcommand{\revtwo}[1]{{#1}}

\begin{abstract}
\revision{
The QSQH theory is extended to all three velocity components taking into account the fluctuations of the direction of the large-scale component of the wall friction. This effect  is found to be significant. It explains the large sensitivity of the fluctuations of longitudinal and spanwise velocities to variations in the Reynolds number in comparison with the sensitivity of the mean velocity, the Reynolds stress, and the wall-normal velocity fluctuations. The analysis shows that the variation of the longitudinal velocity fluctuations with the Reynolds number is dominated by the variation of the amplitude and wall-normal-scale modulation of the universal mean velocity profile by the outer, large-scale, Reynolds-number-dependent motions. The variation of spanwise velocity fluctuations is dominated by the fluctuations of the direction of the large-scale component of the wall friction. The Reynolds number dependence of the other second moments is not dominated by these mechanisms because the mean wall-normal velocity and the mean spanwise velocity are zero. Explicit relationships between the differences in the second moments of velocity in any two high-Reynolds-number near-wall flows were derived. The comparisons gave a satisfactory agreement for the root-mean square of the wall-parallel velocity components in the range of the distances from the wall where modulation by large-scale motions dominates. Relationships between the differences of the constants of the logarithmic law, the shape of the mean velocity profile, and the differences of the second moments of velocity caused by the differences in large-scale motions were derived and estimated quantitatively.
}
\end{abstract}


\newcommand\bydef[1]{\stackrel{\text{def}}#1}
\newcommand\bysym[1]{\stackrel{\text{sym}}#1}
\newcommand\byQSQHu[1]{\stackrel{(\ref{eqn:QSQHonePointu})}#1}
\newcommand\byQSQHv[1]{\stackrel{(\ref{eqn:QSQHonePointv})}#1}
\newcommand\byQSQHw[1]{\stackrel{(\ref{eqn:QSQHonePointw})}#1}
\newcommand\byP[2]{\stackrel{P#1}#2}

\newcommand{\utL}{u_{\tau_L}}
\newcommand{\utLp}{u'_{\tau_L}}
\newcommand{\utLpsq}{u'^2_{\tau_L}}
\newcommand{\mean}[1]{\left\langle{#1}\right\rangle}
\newcommand{\cmean}[1]{\left\langle{#1}\right\rangle_{\utL}}
\newcommand{\der}[2]{\frac{\displaystyle d #1}{\displaystyle d #2}}
\newcommand{\dder}[2]{\frac{\displaystyle d^2 #1}{\displaystyle d {#2}^2}}
\newcommand{\btau}{\boldsymbol\tau}

\newcommand{\enquote}[1]{``#1''}
\newcommand{\commentout}[1]{}

\newcommand{\bx}{\mathbf x} 
\newcommand{\bu}{\mathbf u} 

\newcommand{\tu}{\tilde u}
\newcommand{\tv}{\tilde v}
\newcommand{\tw}{\tilde w}
\newcommand{\tU}{\tilde U}
\newcommand{\ttime}{\tilde t}
\newcommand{\tx}{\tilde x}
\newcommand{\ty}{\tilde y}
\newcommand{\tz}{\tilde z}

\newcommand{\revcom}[1]{}

\section{Introduction} \label{sec:introduction}

The modern understanding of near-wall turbulence has been largely built on the basis of the classical universality hypothesis, according to which as  the Reynolds number, $\Rey$, tends to infinity, the characteristics of the near-wall part of \revision{any turbulent flow}, if expressed in wall units, become independent of  $\Rey$ and other factors such as the pressure gradient \revision{and the wall geometry, provided that the pressure gradient does not lead to separation in the area of interest, and that the wall is smooth. 

\commentout{ CITE?
The law of the wall in turbulent flow
B y P eter B r a d sh a w 1 a nd G eorge P. H u a n g 2

Proc. R. Soc. Lond. A (1995) 451, 165-188
Law of the wall’ is the forceful name for the finding that, with certain assumptions,
the mean velocity U in constant-property turbulent flow near a smooth impermeable
solid surface of negligible curvature can be correlated in terms of the shear stress at
the surface rw, the distance from the surface y, and the fluid properties p (density)
and p (molecular viscosity). 
}

\revision{The wall units are non-dimensional units based on the kinematic viscosity $\nu^*$ and the friction velocity $u_\tau^*=\sqrt{\mean{\tau^*}/\rho^*}$, where $\mean{\tau^*}$ is the mean wall friction and $\rho^*$ is the fluid density. In the present paper the averaging, denoted $\mean{.}$, is defined as an ensemble averaging. It can be understood as the average over a large number of independent experiments, each providing a member of the statistical ensemble. In many practical situations ergodicity can be assumed at least with respect to time and often also with respect to wall-parallel directions. Then the ensemble averaging is equivalent to infinite-time averaging, or averaging in any of the statistically-homogeneous (wall-parallel) directions. The variables expressed in wall units are usually denoted by a superscript ``+'':
\[ 
\bu^+=\bu^*/u_\tau^*, \quad \bx^+ = \bx^* u_\tau^*/\nu^*, \quad t^+ = t^* {u_\tau^*}^2/\nu^*,
\]
where $\bu^*, \bx^*$, and $t^*$ are the dimensional velocity, radius-vector, and time respectively.
}

The classical universality hypothesis can be expressed by not including $\Rey$ into the arguments of the expression
\begin{equation}\label{eqn:CUP}
\bu^+=\bu^+(t^+,\bx^+).
\end{equation}
This implies that the right-hand side is independent of $\Rey$ and other flow parameters. Here, independence is understood in a statistical sense. This means that all statistics  of $\bu^+(t^+,\bx^+)$, that is averages of all function(al)s of $\bu^+(t^+,\bx^+)$, are independent of $\Rey$.
}%

Research in the last two decades has shown that the classical universality hypothesis is at least inaccurate. 
High-$\Rey$ turbulent near-wall flows exhibit two peaks in the turbulent kinetic energy profile \citep[see e.g.][]{smits11}. The first peak is located in the buffer layer, while the distance from the wall to the second, \enquote{outer}, peak varies with $\Rey$, when measured in wall units.
The outer peak is caused by turbulent fluctuations at characteristic length scales that are significantly larger than those causing the inner peak \citep{hutchins07}. 
These outer large-scale motions and the inner small-scale motions located in the near-wall region interact~\citep{rao71,mathis09a,gana12}.
\citet{marusic10sci} proposed the widely known empirical relation describing this interaction.

 Since the large-scale motions are not $\Rey$-independent if expressed in wall units, the scale interaction invalidates the classical universality hypothesis. The recently developed Quasi-Steady-Quasi-Homogeneous (QSQH) theory~\citep{ChernyshenkoEtAl:2012:arxiv,Zhang16,Chernyshenko_FDR2019} remedies this. In essence, the QSQH hypothesis states that the small-scale near-wall motions adapt to the large-scale fluctuations  of the wall shear stress.

Results agreeing to various degree  with the QSQH theory were obtained in 
\citep{ChernyshenkoEtAl:2012:arxiv, Agostini:2014:PoF,Agostini2018, Zhang16, Baars.2016.PhysRevFluids.1.054406,Baars.2017.rsta,agostini2017multi,howland2018dependence,Chernyshenko_FDR2019,Zhangthesis, agostini2019connection,  lozierexperimental}. 
The theory performed better for higher $\Rey$ and smaller distances to the wall. In most cases, significant deviations occurred only for $y^+>70$ wall units.
\revision{ It was not tested for $\Rey_\tau<950$, since at such small $\Rey_\tau$ there are virtually no large-scale motions. The majority of checks were done on the basis of direct numerical simulations of channel flows. However, the empirical relation of \citet{marusic10sci} coincides with the main term of the Taylor expansion of the QSQH theory in terms of the amplitude of the large-scale fluctuations~\citep{ChernyshenkoEtAl:2012:arxiv}. Therefore, the numerous confirmations of this empirical relation by experiments in boundary layers also support the QSQH theory.

The majority of the comparisons considered only the longitudinal component of the velocity.
}
Recently, \citet{agostini_leschziner_2019} attempted to extend the QSQH theory, stated originally for the longitudinal velocity only,  to the spanwise and wall-normal velocity components, and, for $\Rey_\tau=1000$, found that while for the wall distance $y^+<70$ the behaviour of the longitudinal, $u$, and wall-normal, $v$, velocity components was in approximate agreement with their extension of the QSQH theory, the behaviour of the spanwise velocity component $w$ was in a noticeable disagreement even in the close vicinity of the wall.  They attributed this discrepancy to the effect of the wall-normal component of the large-scale motions, but also mentioned that taking into account that the wall friction is a vector, as suggested by ~\citet{Zhang16}, might improve the performance of the theory.
The observed discrepancy is intriguing because velocity components are related by the continuity equation and, at a first glance, an agreement in two components should lead to an agreement in the third component.

Experimental studies concentrate mostly on the longitudinal components of the velocity, but we can mention the works \citep{baidya2012} and \citep{talluru2014amplitude} containing observations relevant to the present study.
Research involving comparisons for all three velocity components is likely to be continuing. It is therefore timely to provide the formulation of the QSQH theory taking into account the fluctuation of the direction of the large-scale wall friction. This is the main goal of the present paper. 
 First results within the extended theory and a brief discussion are also presented.

\revtwo{The rest of the paper is organised in the following way. Section~\ref{sec:Extension} describes the theory, its basic relations and typical derivation techniques. It is stated for all three velocity components and in the form applicable for the case when the fluctuation of the direction of the large-scale wall friction is not neglected, which is the novel contribution. Where the spanwise and wall-normal velocity components and the direction of large-scale wall friction are not explicitly involved, as for example, in subsections~\ref{sec:Large-scale motions} and~\ref{sec:Non-dimensional units}, this coincides with that in the previous works~\citep{ChernyshenkoEtAl:2012:arxiv,Zhang16}, and it is provided here for completeness. Section~\ref{sec:moments} is devoted to the theoretical predictions for the dependence of mean velocity and second moments on the Reynolds number. Section 4 is about the non-universality of  the constants of the logarithmic law. A brief Section~\ref{sec:OnAgostiniLeschziner} describes an attempt to analyse the results of the recent work of~\citet{agostini_leschziner_2019}. It is followed by a discussion and conclusions.

}

\section{Extension of the QSQH theory to all velocity components}\label{sec:Extension}

\revcom{R2:  I would suggest a
slightly expanded introduction to the theory and notation (opening paragraphs
of section 2).}
 
\subsection{Large-scale motions}\label{sec:Large-scale motions}

Giving a rigorous definition of large scale motions is a challenge. One of the most common definitions uses a Fourier cut-off filter in time, or one of homogeneous spatial directions, or a combination of such filters. The alternatives include  the bi-dimensional version of the empirical mode decomposition \citep{HuangEMD1998} first applied in the present context in \citep{Agostini:2014:PoF} and the decomposition based on the hierarchical random additive process \citep{YangEtAl2016PRF}. 
 Each of these methods requires additional decisions to be made to define the filter uniquely. The QSQH theory \citep{Zhang16} circumvents this difficulty by specifying the properties of the large-scale filter that are required for the theory to be valid. Then the theory is built without defining the filter itself. To apply the theory to a particular flow one can then select the filter satisfying these properties at least approximately. An extensive investigation of filter selection was done by \cite{Zhangthesis}. Alternatively, one can derive useful relationships on the basis only of the assumption that such a filter exists, without specifying the filter explicitly. This approach will be followed in this paper. The reasons why this is at all possible are discussed in \citep{Chernyshenko_FDR2019}, but this issue is far outside the scope of the present paper.

\revision{
A large-scale filter is an operator that if applied to any scalar or vector-valued function of time and spatial coordinates gives the large-scale component of that function. We will denote the large-scale component with a subscript ``$L$''. Thus, $f_L(t,\bx)=\LSF f(t,\bx)$, where $\LSF$ is the large-scale filter. 

The QSQH theory assumes that  $\LSF$ has the following properties. For any constants $a$ and $b$ and functions $f (t,\bx)$ and $g(t,\bx)$:
\smallskip

\noindent P1. Linearity,
\begin{equation}\label{eqn:Li}
\LSF (af + bg) = a\LSF f + b\LSF g,
\end{equation}
P2.  Invariance of averages: the averaged variables are large-scale,
\begin{equation}\label{eqn:Lii} 
\LSF \mean{f } = \mean{f },
\end{equation}
P3. Projection property: the large-scale filter does not change an already large-scale-filtered
function,
\begin{equation}\label{eqn:Liii}  
\LSF \LSF f = \LSF f,
\end{equation}
P4. Commuting with averaging,  
\begin{equation}\label{eqn:Liv} 
\mean{\LSF f } = \LSF \mean{f }, 
\end{equation}
P5.  Scale-separation property: applying the large-scale filter to any function of $t$, $\bx$, and other
arguments that are large-scale-filtered variables is equivalent to averaging with the large-scale arguments held constant:
\begin{equation}\label{eqn:Lv} 
\LSF f (t,\bx, Lg_1, \dots, \LSF g_n) = \mean{f (t,\bx,\xi_1, . . . ,\xi_n )}_{\xi_1=\LSF g_1,\dots,\xi_n=\LSF g_n}.
\end{equation}
The fifth property expresses the reason why the QSQH theory might describe the scale interaction in near-wall turbulence. Property 5 will be satisfied to the extent to which the characteristic dimensions of the large scales are much larger than the characteristic dimensions of small scales.  

}

\subsection{\revision{Non-dimensional units}}\label{sec:Non-dimensional units}
\revision{
Wall friction $\btau^*$ and its large-scale component $\btau_L^*=\LSF \btau^*$ are vectors parallel to the wall. Similar to the friction velocity, we introduce the large-scale friction velocity
\begin{equation}\label{eqn:utldef}
\utL^*=\sqrt{|\btau_L^*|/\rho^*}.
\end{equation} 
Note that unlike the friction velocity,  $\utL^*$ is a function of time and the wall-parallel coordinates, and, where the statistical interpretation is used, it is a random function.
In the rest of the paper all quantities will be made non-dimensional using the  average large-scale friction velocity $\mean{\utL^*}$. Namely, the non-dimensional velocity, coordinates, and time are
\begin{equation}\label{eqn:nondim}
\bu=\bu^*/\mean{\utL^*}, \quad \bx = \bx^* \mean{\utL^*}/\nu^*, \quad t = t^* {\mean{\utL^*}}^2/\nu^*.
\end{equation}
where starred variables are dimensional. 
In these units the fluctuation of the large-scale friction velocity  $\utLp=\utL-1$. The friction velocity is $
u_\tau=\sqrt{\mean{\tau}}=\sqrt{\mean{\tau_L}}=\sqrt{\mean{\utL^2}}=\sqrt{1+\mean{\utLpsq}}$, where we used that $\mean{\tau}=\mean{\tau_L}$ by a combination of Properties 2 and 4.

The variables in the usual wall units and the non-dimensional units thus introduced 
 are therefore related as
 \begin{equation}\label{eqn:wallunits}
 \bu^+=\bu/u_\tau=\bu/\sqrt{1+\mean{\utLpsq}},\quad \bx^+=\bx\sqrt{1+\mean{\utLpsq}}, \quad t^+=t\left(1+\mean{\utLpsq}\right). 
 \end{equation}
 They are very close quantitatively, at least for moderate $\Rey$, when the fluctuations of the magnitude of large-scale motions are small. 

}

\subsection{\revision{Formulation of the QSQH hypothesis}}

 If the large-scale component $\btau_L$ were independent of time and the wall-parallel coordinates, while the small-scale component had zero time average, one could use wall units based on $\btau_L$, which would coincide with the mean wall friction, and expect that near the wall the flow statistics expressed in thus-defined wall units would be (approximately) independent of $\Rey$ and other factors, including the dimensional value of $\btau_L^*.$  The QSQH theory assumes that the same remains true even when $\btau_L$ is dependent on time and the wall-parallel coordinates. The physical meaning of this hypothesis is therefore that the near-wall flow adjusts to the large-scale wall friction as if it were constant in time and homogeneous in space. 

\begin{figure}
\setlength{\unitlength}{1 ex}
	\centering\begin{picture}(70,30)
\put(0,0){\includegraphics[width=70 ex, clip, trim = 0 200 0 100]{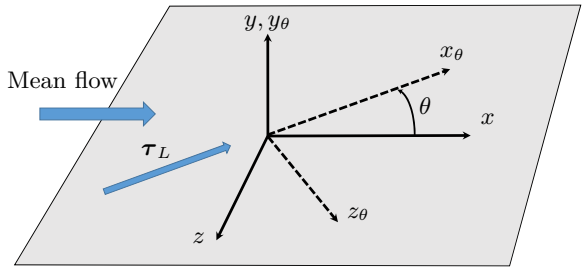}}
\put (47,16.5){$\theta$}
\put(53,15.5){$x$}
\put(30,25){$y,y_\theta$}
\put(25,4){$z$}
\put(49,22){$x_\theta$}
\put(40,6.5){$z_\theta$}
\put (20,12.5){$\btau_L$}
\put (7,19){Mean flow}
\end{picture}
	\caption{The coordinate systems: $x$ is in the direction of the mean flow, $x_\theta$ is in the direction of the large-scale component of the wall friction, $y$ and $y_\theta$ coincide, being in the wall-normal direction, $z$ and $z_\theta$ are in the wall plane and perpendicular to $x$ and $x_\theta$ respectively. The angle $\theta$ is measured clockwise if looked at from above the wall.	\label{fig:coord}}
	\end{figure}


 The large-scale component of the wall friction $\btau_L(t,x,z)$ is a vector.  
A local coordinate system with an axis along the direction of  $\btau_L(t,x,z)$ shown in  figure~\ref{fig:coord} and the coordinate system $x,y,z$ aligned with the mean flow direction are related by the formulae
\begin{align}
x_\theta=x\cos\theta -z \sin\theta,\quad
y_\theta=y,\quad
z_\theta=x\sin\theta+z\cos\theta, \label{eqn:QSQHxyz}
\\
u=u_\theta\cos\theta +w_\theta \sin\theta,\quad
v=v_\theta,\quad
w=-u_\theta\sin\theta+w_\theta\cos\theta,\label{eqn:QSQHw}
\end{align}
where $\theta$ is the angle between the direction of the mean wall friction and the large-scale wall friction. 

The main relationships of the extended QSQH theory are
\begin{align}
u_\theta&(t,x_\theta,y_\theta,z_\theta)=\utL\tilde u(\tilde t,\tx,\ty,\tz),\label{eqn:QSQHutheta}\\
v_\theta&(t,x_\theta,y_\theta,z_\theta)=\utL\tilde v(\tilde t,\tx,\ty,\tz),\label{eqn:QSQHvtheta}\\
w_\theta&(t,x_\theta,y_\theta,z_\theta,)=\utL\tilde w(\tilde t,\tx,\ty,\tz),\label{eqn:QSQHwtheta}\\
\tilde t=t\utL^2&,\ \ \tx=x_\theta\utL,\ \ \ty=y_\theta\utL,\ \ \tz=z_\theta\utL,       \label{eqn:tilde_variables}\\
P[\utL,\theta,&\tilde u, \tilde v, \tilde w,\Rey]=P_\text{o}[\utL,\theta,\Rey]P_\text{u}[\tilde u, \tilde v, \tilde w],\label{eqn:PDFs}  
\end{align}
\revision{%
where $P$ is the joint probability density functional of the functions $\utL(\cdot)$, $\theta(\cdot)$, $\tilde u(\cdot)$, $\tilde v(\cdot),$ $ \tilde w(\cdot)$ and  $P_\text{o}$ and $P_\text{u}$ are marginal probability density functionals of  $\utL(.)$, $\theta(.)$ and $\tilde u(.),$ $ \tilde v(.)$, $\tilde w(.)$ respectively. 
Similar to the expression of the classical universality hypothesis~(\ref{eqn:CUP}), the absence of $\Rey$ from the arguments of $P_\text{u}$  is meant to indicate that the functions $\tilde u(.)$, $\tilde v(.)$, and $\tilde w(.)$ are universal in the sense that their statistical properties are independent of  $\Rey$ and of other global flow properties.

Equation (\ref{eqn:PDFs}) means that the set of functions $\utL(\cdot)$, $\theta(\cdot)$ and the set of functions $\tilde u(\cdot),$ $ \tilde v(\cdot)$, $\tilde w(\cdot)$ are statistically independent. To clarify,
 this for example implies that for any non-random constant numbers $\xi_1,\xi_2,\xi_3,\xi_4,t,x$, and $z$ the value of $\tilde u(\xi_1,\xi_2,\xi_3,\xi_4)$ and the value of $\utL(t,x,z)$ are statistically independent, which in turn implies that their correlation is zero.  Note however that for constant $t,x,y,z$  the values of $\tilde u(\tilde t,\tx,\ty,\tz)$ and   $\utL(t,x,z)$ are not statistically independent, since in this case $\tilde t,\tx,\ty,\tz$ depend on $\utL$. 
Note also that though~(\ref{eqn:QSQHutheta}-\ref{eqn:tilde_variables}) do not involve $\theta$ explicitly, the statistics of $u,$ $v,$ and $w$ depend on the statistics of $\theta$ since it enters  (\ref{eqn:QSQHxyz},\ref{eqn:QSQHw}).
}

If $\utL$ and $\theta$ were not fluctuating then $\utL$ would be equal to 1 due to our choice of the non-dimensional variables, the averaged large-scale friction would be equal to the mean friction,  our non-dimensional units would coincide with the wall units,   $\tilde u, \tilde v, \tilde w$ would be the velocity components in wall units $u^+, v^+, w^+$, and $\tilde t,\tx,\ty,\tz$ would be the time and coordinates in wall units $t^+, x^+,y^+,z^+$. Equations (\ref{eqn:QSQHutheta}-\ref{eqn:tilde_variables}) are the QSQH theory refinement  suggested, although not explicitly formulated, by \cite{Zhang16}.
 If only the magnitude of the wall friction but not its direction fluctuates, that is $\theta=0$, then (\ref{eqn:QSQHutheta}) coincides with (5) in \cite{ChernyshenkoEtAl:2012:arxiv}, and (\ref{eqn:QSQHvtheta},\ref{eqn:QSQHwtheta}) are the assumptions implied by~\citet{agostini_leschziner_2019}.

\revision{The variables marked with tilde are the QSQH analogue of the variables marked with plus in the classical universality paradigm. These 'tilde' variables are a kind of `plus' variables, but built on the fluctuating in magnitude and direction large-scale component of skin friction instead of constant mean skin friction.

Note also that as it follows from their definition and the filter Property 5 (\ref{eqn:Lv}), $\utL$ and $\theta$ are large-scale functions, so that $\LSF\utL=\utL$ and $\LSF\theta=\theta$.
}

\revision{
The QSQH theory can be thought of as consisting of three parts. The first part 
is a rigorous mathematical theory of the relations between vector-valued functions $\bu(t,x,y,z)$ and $\btau_L(t,x,z)$,  functions $\utL(t,x,z)=\sqrt{|\btau_L(t,x,z)}|$ and $\theta(t,x,z)$, and an operator $\LSF$ satisfying the relations (\ref{eqn:Li}-\ref{eqn:Lv}, \ref{eqn:QSQHxyz}-\ref{eqn:QSQHw}, \ref{eqn:QSQHutheta}-\ref{eqn:PDFs}).  The second part, a physical hypothesis of quasi-steadiness, is the hypothesis that there exists such an operator $\LSF$ that the velocity field of a real fluid flow satisfies, at least approximately, the above conditions.
The third part, a physical hypothesis of universality, is the hypothesis that the statistical properties of the  functions $\tilde u(\cdot,\cdot,\cdot)$, $\tilde v(\cdot,\cdot,\cdot)$, and $\tilde w(\cdot,\cdot,\cdot)$   are at least approximately the same for all real flows with sufficiently large $\Rey$ independently of other large-scale factors such as the steady pressure gradient or smooth wall geometry. With both physical hypotheses accepted, the QSQH theory allows to express the statistical properties of the total velocity field in terms of the statistical properties of $\utL$ and $\theta$, which depend on $\Rey$ and other large-scale factors, and the universal statistical properties of  $\tilde u(\cdot,\cdot,\cdot)$, $\tilde v(\cdot,\cdot,\cdot)$, and $\tilde w(\cdot,\cdot,\cdot)$. 
}


\subsection{Simplifications and basic relations}

If only the statistics that are one-point in wall-parallel directions and time are considered,  shifting the coordinate origin to the point and time of interest makes $t=x=x_\theta=z=z_\theta=0$. Then with zero arguments omitted (\ref{eqn:QSQHxyz}-\ref{eqn:QSQHwtheta}) reduce to
\begin{align}
u/\utL&=\tilde u(y\utL)\cos\theta +\tilde w(y\utL) \sin\theta,\label{eqn:QSQHonePointu}\\
v/\utL&=\tilde v(y\utL),\label{eqn:QSQHonePointv}\\
w/\utL&=-\tilde u(y\utL)\sin\theta+\tilde w(y\utL)\cos\theta.\label{eqn:QSQHonePointw}
\end{align}
Further considerations in the present paper will refer to this case. Once again, we use ensemble averaging, with the understanding that with the ergodicity assumption the ensemble averages are equal to time and homogeneous direction averages, to which the obtained results will therefore also apply. 

\revision{
When the large-scale wall friction does not fluctuate and is the same everywhere on the wall, $\theta=0$, $\utL=1$, $x=\tx$, $y=\ty$, $z=\tz$, $u=\tu$, $v=\tv$, and  $w=\tw$. A natural assumption that the mean velocity $(U,V,W)$ of such a flow is parallel to the wall friction and satisfies continuity and impermeability conditions means that  $U(y)=\tU(\ty)=\mean{\tu(\ty)}\ne 0$ but  
\begin{equation}\label{eqn:ZeroMeans}
V(y)=\tilde V(\ty)=\mean{\tv(y)}=W(y)=\tilde W(\ty)=\mean{\tw(\ty)}=0. 
\end{equation}
}
\revision{
Applying the large-scale filter to the Newton friction law, which in dimensional form is $\tau^*_x/\rho =\nu^*\left.\pd {u^*}{y^*}\right|_{y^*=0}$, $\tau^*_z/\rho =\nu^*\left.\pd {w^*}{y^*}\right|_{y^*=0}$ and combining the result with the definition of $\utL^*$ (\ref{eqn:utldef}) and the expressions  (\ref{eqn:LS}) in subsection~\ref{sec:commonders} gives $d\tU/d\ty=1$ at $\ty=0$. This is fully analogous to the well-known relation $dU^+/dy^+=1$ at the wall.
}

\revision{
\subsection{Common derivation techniques and mean velocity}\label{sec:commonders}

Typical application of the QSQH theory involves expressing the flow statistics in terms of the statistics of the universal velocity $(\tu, \tv, \tw)$ and the statistics of $\utL$ and $\theta.$ Several techniques proved to be particularly useful. The typical derivation of the expression for  an average, for example $\mean{f}$, starts with applying  filter Properties 2 and 4,  (\ref{eqn:Lii}) and (\ref{eqn:Liv}), to get $\mean f=\mean{\LSF f}$. Then Property 5, (\ref{eqn:Lv}), is used to express $\LSF f$ in terms of its average with a part of the arguments assumed to be constant. This intermediate average will typically be given a notation, for example, $\tilde F$, and will still be a function of those arguments of $f$ that are large-scale quantities and were, therefore, held constant when Property 5 was applied. Typically, $\tilde F$ would be expressed in terms of the averages of the universal functions. A frequent simplification follow from the left-right symmetry considerations: for example $\mean {\sin\theta}=0$ and $\mean{\utL\sin\theta}=0$ because the negative and positive values of $\theta$ are equally probable. Another typical simplification follows from statistical independence, expressed by (\ref{eqn:PDFs}), of the large-scale skin friction, represented by $\utL$ and $\theta$, and the universal functions. The derivations based on these techniques, which we will call the QSQH derivations, are routine, often lengthy but, fortunately, often lead to compact outcomes.  For example, for the mean velocity the QSQH derivation (see Appendix) gives
}
\revision{%
\begin{equation}\label{eqn:U}
U(y)=
 \mean{\utL \tilde U(y\utL)\cos\theta},
\end{equation}
}
where the mean universal velocity is defined as
$
\tU(\xi)\bydef=\mean{\tu(\xi)}_{\xi=\text{const}}
$.

\revision{
For  $\Rey$ achievable so far in direct numerical simulations the amplitude of the fluctuation of large-scale motions is small.  Constructing a truncated Taylor expansion in powers of  $\theta$ and $\utLp=\utL-1$  or their statistical moments  is another typical technique. Note that the coefficients in the Taylor expansions obtained within the QSQH theory can sometimes be quite large~\citep{Zhang16}. Hence, caution is required in determining the sufficient number of terms. For the mean velocity the Taylor expansion gives
\begin{multline*}
U(y)= \mean{(1+\utLp) \tU(y+y\utLp)\cos\theta}\\
=\mean{(1+\utLp)\left(\tU(y)+\der\tU y y\utLp+\dder\tU y \frac{(y\utLp)^2}2+\dots\right)\left(1-\frac{\theta^2}2+\dots\right)}\\
=\tU(y)+\mean{\utLp}\tU(y)+\frac12\mean{\utLpsq}\left(2y\der\tU y +y^2\dder\tU y\right)-\frac12\mean{\theta^2}\tU+\dots.
\end{multline*}
}
Since $\mean{\utLp}=0$, a simple further transformation gives

\revcom{R2: The Taylor series expansions could be covered in a little more
detail, for example I struggled for a while to understand exactly where (2.10)
came from.}

\begin{equation}\label{eqn:UTT}
U(y)=\tilde U(y)+\frac12\mean{\utLpsq} \der\ y y^2\der{\tU}y - \frac12\mean{\theta^2}\tU(y)+\dots.
\end{equation}

\subsection{Large-scale and small-scale motions within the QSQH theory}

For the large-scale velocity components, defined as $u_L(y)=\LSF u(y)$, $v_L(y)=\LSF v(y)$, and $w_L(y)=\LSF w(y)$, the QSQH derivation (see Appendix) gives:
\begin{equation}\label{eqn:LS}
\begin{aligned}
u_L&=\utL \tilde U(y\utL)\cos\theta,\\
v_L&=0,\\
w_L&=-\utL\tU(y\utL)\sin\theta.
\end{aligned}
\end{equation}
Hence, the large-scale component of the velocity has a unidirectional velocity profile, the direction of which coincides with the direction of the large-scale component of the wall friction. The shape of this profile is the same as the shape of the universal mean velocity profile stretched by a factor of $1/\utL$ in the wall-normal direction and of the magnitude of $\utL$ times greater than the magnitude of the universal mean profile.     
The small-scale velocity components are defined as $u_S=u-u_L$, $v_S=v-v_L$, and $w_S=w-u_L$. The QSQH derivation (see Appendix) gives that the correlations of  large and small components are zero:
\begin{equation}\label{eqn:LScorr}
\mean{u'_Lu'_S}=\mean{u'_Lv'_S}=\mean{u'_Lw'_S}=\mean{w'_Lu'_S}=\mean{w'_Lv'_S}=\mean{w'_Lw'_S}=0.
\end{equation}

This completes a brief description of the theory including the properties of the filter, the ideas of the main techniques used in the derivations, and a few basic results. Now we will show that taking into account the fluctuations of the direction of large-scale friction makes a difference.

\section{Dependence of mean velocity and second moments on $\Rey$}\label{sec:moments}

\subsection{Sensitivity of mean velocity and the second moments to changes in $\Rey$}\label{sec:momentssensitivity}

\begin{figure}
\setlength{\unitlength}{1 ex}
	\centering\begin{picture}(90,28)
\put(0,2){\includegraphics[height=25 ex]{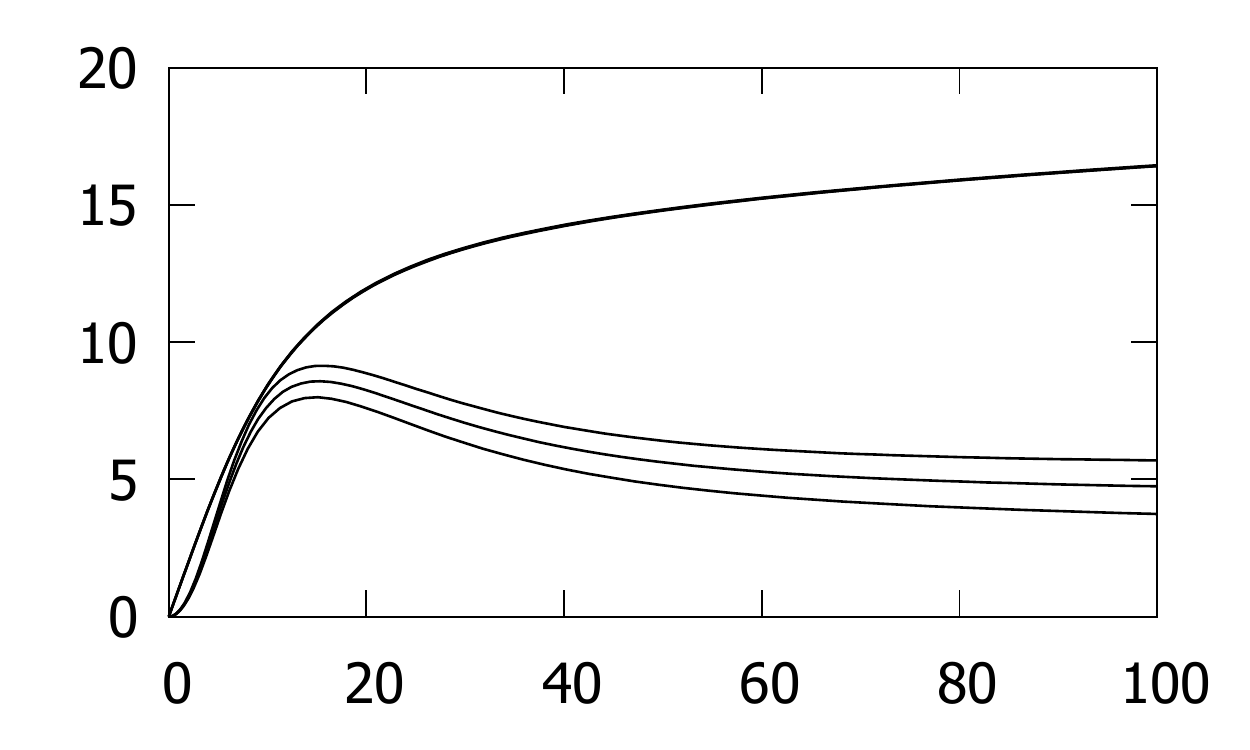}}
\put(45,2){\includegraphics[height =25ex]{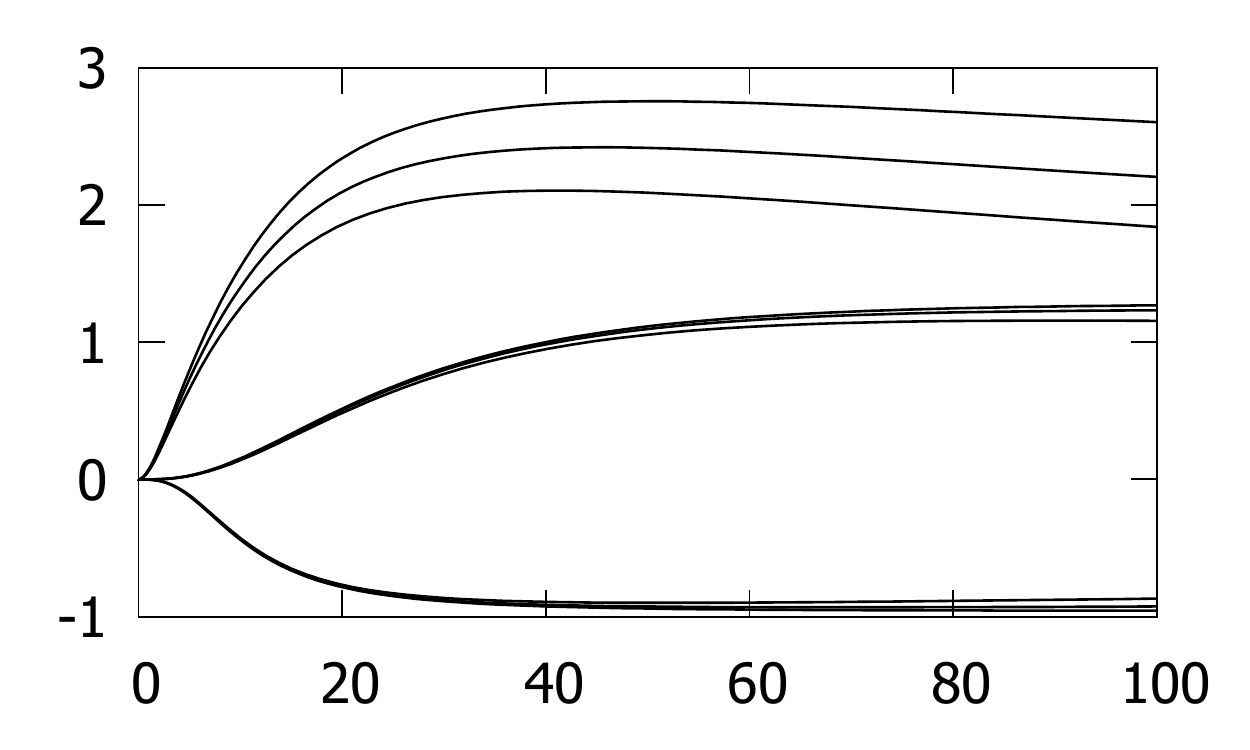}}
\put(25,0.5){$y^+$}
\put(70,0.5){$y^+$}
\put (23,21){$U$}

\put (29,13){$u_\text{rms}^2$}
\put(16,9.5){$\Rey_\tau$ \vector(1,2){2}}

\put(50,22){$w_\text{rms}^2$}
\put(70,18){$\Rey_\tau$ \vector(1,1){6}}

\put(75,15){\vector(1,1){3}}
\put(77,13){$v_\text{rms}^2$}

\put(75,8){$\tau_R$}
\put(55,10){\vector(-1,-1){3} $\Rey_\tau$}
\end{picture}
	\caption{The DNS data for $\Rey_\tau=950$, $2000$ and $5200$ showing that $u_\text{rms}^2$ and  $w_\text{rms}^2$ vary with $\Rey_\tau$ significantly more than $U$, $v_\text{rms}^2$ and $\tau_R=\mean{u'v'}$. The three curves for $U$ overlap. Note the difference in the scales of the ordinate axes in the plots.\label{fig:moments}}
	\end{figure}
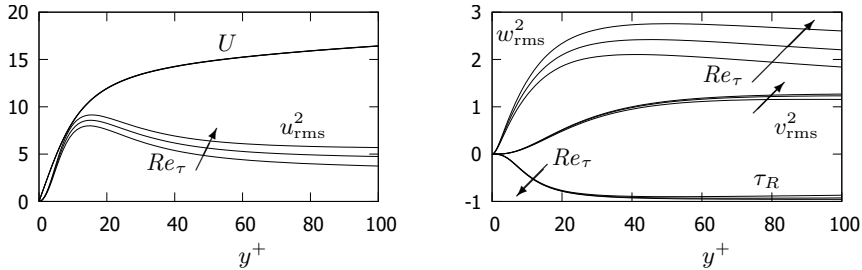

It is known but remained unexplained so far that both in absolute and in relative terms the variation of $u_\text{rms}^2=\mean{u'^2}$ and $w_\text{rms}^2=\mean{w'^2}$ with $\Rey_\tau$ is much greater than the variation of  $U(y),$   $v_\text{rms}^2=\mean{v'^2}$, 
and the Reynolds stress 
$\tau_R=\mean{u'v'}$,  see figure~\ref{fig:moments}. This figure uses the direct numerical simulations (DNS) of a plane channel flow data  available on the web at turbulence.oden.utexas.edu. The DNS are described in \citep{hoyas06} and \citep{LeeMoserJFM2015}. For the analysis, we fitted  the velocity average and its nonzero second moments with explicit expressions. The fit is described in the Supplement.

We introduce the fluctuation of the universal  streamwise velocity $\tu'(\ty)=\tu(\ty)-\tU(\ty)$,  the universal second moments $\tu_\text{rms}^2(\ty)=\mean{{{\tu'\!\,}}^2(\ty)}_{\ty=\text{const}}$, $\tv_\text{rms}^2(\ty)=\mean{\tv(\ty)^2}_{\ty=\text{const}}$, $\tw_\text{rms}^2(\ty)=\mean{\tw(\ty)^2}_{\ty=\text{const}}$, and the universal mean Reynolds stress $\tilde \tau_R(\ty)=\mean{\tu'(\ty)\tv(\ty)}_{\ty=\text{const}}$. 
Applying the same technique as the technique used to derive (\ref{eqn:U}) gives (see Appendix)
\begin{multline}\label{eqn:urmsF}
u_\text{rms}^2(y)=\mean{(u-U)^2}=\mean{\left(\utL\tU(y\utL)\cos\theta\right)^2}-U^2+\\
\mean{\utL^2\left(\tu_\text{rms}^2(y\utL)\cos^2\theta+
\tw_\text{rms}^2(y\utL)\sin^2\theta\right)},
\end{multline}

\begin{equation}\label{eqn:vrmsF}
v_\text{rms}^2(y)
=\mean{\utL^2\tv_\text{rms}^2(y\utL)},
\end{equation}
\begin{multline}\label{eqn:wrmsF}
w_\text{rms}^2(y)
=\mean{w^2(y)}=\mean{\left(\utL\tU(y\utL)\sin\theta\right)^2}+\\
\mean{\utL^2\left(\tw_\text{rms}^2(y\utL)\cos^2\theta+
\tu_\text{rms}^2(y\utL)\sin^2\theta\right)},
\end{multline}
\begin{equation}\label{eqn:uvF}
\mean{u'(y)v'(y)}=\mean{\utL^2\tilde \tau_R(y\utL)\cos\theta}.
\end{equation}

The Taylor expansions of  (\ref{eqn:urmsF}-\ref{eqn:uvF}) in powers of  $\theta$ and $\utLp=\utL-1$ are

\begin{alignat}{2}
u_\text{rms}^2(y)&=\mean{\utLpsq}\left(\der{y\tU}y\right)^2\!\!\!&{}+{}&\tu_\text{rms}^2(y)+\frac{\mean{\utLpsq}}2 \dder{y^2\tu_\text{rms}^2}y+
\mean{\theta^2}\left(\tw_\text{rms}^2-\tu_\text{rms}^2\right)+\dots, \label{eqn:urmsT}\\
v_\text{rms}^2(y)&= &\hphantom{{}+{}}&\tv_\text{rms}^2(y)+\frac12\mean{\utLpsq}\dder{y^2\tv_\text{rms}^2}y+\dots, \label{eqn:vrms}\\
w_\text{rms}^2(y)&=\qquad\mean{\theta^2}\tU^2&{}+{}&\tw_\text{rms}^2(y)+\frac12\mean{\utLpsq}\dder{y^2\tw_\text{rms}^2}y+\mean{\theta^2}\left(\tu_\text{rms}^2-\tw_\text{rms}^2\right)+\dots,\label{eqn:wrms}\\
\mean{u'v'}(y)& = &\hphantom{{}+{}}&\tilde \tau_R(y)+\frac12\mean{\utLpsq}\dder{y^2\tilde \tau_R}y-\frac12\mean{\theta^2}\tilde \tau_R +\dots .\label{eqn:uv} 
\end{alignat}
\revision{Note that for any given $\utL$ negative and positive values of $\theta$  are equally probable (this is a left-right symmetry, similar for example to why the average of the fluctuation of the spanwise velocity is zero). Therefore,  $<\utLp\theta>=0$.
}

In the right-hand sides of  (\ref{eqn:U}) and (\ref{eqn:urmsT}-\ref{eqn:uv}) only $\mean{\utLpsq}$ and $\mean{\theta^2}$ vary with $\Rey.$ 
 In (\ref{eqn:U}), $\mean{\utLpsq}$ and $\mean{\theta^2}$ are multiplied by a factor of the same order of magnitude as $U(y)$ itself. Since variation of $U(y)$ with $\Rey$ is very small, the corresponding variation  of $\mean{\utLpsq}$ and $\mean{\theta^2}$ should also be small. Indeed, for $\Rey_\tau$ in the range being considered (between 950 and 5200), the amplitude of large-scale motions is small. For example, for the filter used in \citep{Zhang16},   $ \mean{\utLpsq}= 0.004364 
$ for $\Rey_\tau=1000$, and for the filter used in \citep{Zhangthesis}, $\mean{\utLpsq}= 0.0031$ for $\Rey_\tau=950$ and $ \mean{\utLpsq}= 0.0064$ for $\Rey_\tau=4200$. This allows to conclude that $\tU(y)\approx U(y)$, and by the order of magnitude 
$\tu_\text{rms}^2\sim  u_\text{rms}^2$, $\tv_\text{rms}^2\sim  v_\text{rms}^2$, $\tw_\text{rms}^2\sim  w_\text{rms}^2$, and 
$\tilde \tau_R\sim  \mean{u'v'}(y)$. 
Using this for estimating the terms in (\ref{eqn:U}) and (\ref{eqn:urmsT}-\ref{eqn:uv}) shows that $u_\text{rms}^2(y)$ and $w_\text{rms}^2(y)$ stand out. In (\ref{eqn:urmsT}) $\mean{\utLpsq}$ is multiplied by $(dy\tU/dy)^2$ and in (\ref{eqn:wrms})  $\mean{\theta^2}$ is multiplied by $\tU^2$. 
Outside the close vicinity of the wall these factors are large. At $y^+=40$ for example,  $(dy\tU/dy)^2\approx 285,$ which is almost 50 times greater than $u^2_\text{rms}\approx 6$, and $\tU^2\approx203,$ which is almost 85 times greater than $w^2_\text{rms}\approx 2.4.$  There are no such large terms in (\ref{eqn:U}), (\ref{eqn:vrms}), and (\ref{eqn:uv}). 
The term $\mean{\utLpsq}(d{y\tU}/dy)^2$ describes the combined action of the amplitude modulation and modulation of the wall-normal length scale of the universal mean profile, which is a QSQH mechanism described in \citep{ChernyshenkoEtAl:2012:arxiv,Zhang16}. 
The term $\mean{\theta^2}\tU^2$ describes the turning of the universal mean profile  to follow the fluctuating direction of the large-scale wall friction. Such large terms are not present in the expressions for other quantities.
This explains why $u_\text{rms}^2$ and $w_\text{rms}^2$ vary more  with $\Rey$ than
$U$,  $w_\text{rms}^2$ and $\mean{u'v'}$.
     
\revision{This result allows to formulate an important observation. The effect of the fluctuation of the direction of the large-scale motion on the statistics of the spanwise velocity is substantial. To confirm these theoretical results, comparisons should be made.}

\subsection{The profiles of the change in the second moments with $\Rey$} 

The values of $\mean{\utLpsq}$ and $\mean{\theta^2}$ and the statistics of the universal velocity depend on the definition of the large-scale filter. Luckily, since  $\mean{\utLpsq}$ and $\mean{\theta^2}$ are small, $\tU$ is known quite accurately, and the second moments can be estimated approximately by taking  $\tu_\text{rms}^2\approx u_\text{rms}^2$, $\tv_\text{rms}^2\approx  v_\text{rms}^2$, $\tw_\text{rms}^2\approx w_\text{rms}^2$, and 
$\tilde \tau_R\approx  \mean{u'v'}$.  
 Moreover, the dominance of the terms $\mean{\theta^2}\tU^2$ and $\mean{\utLpsq}(d{y\tU}/dy)^2$ allows to express the shape of the increase in $u_\text{rms}^2$ and $w_\text{rms}^2$  with $\Rey$ as a function of the wall distance in terms of the shape of the mean velocity profile only.

\revcom{R1: The complete derivation of equations (3.5)-(3.8) should be detailed / why there are no $< u0
tL \theta >$ For any given $\utL$ negative and positive $\theta$  are equally probable (this is a left-right symmetry, similar for example to why the average of the fluctuation of the spanwise velocity is zero). Therefore,  $<\utLp\theta>=0$. }

Subtracting  (\ref{eqn:UTT},\ref{eqn:urmsT}-\ref{eqn:uv}) written down for one value of $\Rey$ from the same equations for another value of $\Rey$ gives the increments of the mean velocity and the second order moments as $\Rey$ changes,  expressed in terms of the increments of $\mean{\utLpsq}$ and  $\mean{\theta^2}$:

\begin{equation}\label{eqn:dU}
\Delta U(y)=\frac12\DELT\mean{\utLpsq} \der\ y y^2\der{\tU}y - \frac12\DELT\mean{\theta^2}\tU(y)+\dots,
\end{equation}
\begin{equation}\label{eqn:durms}
\Delta u_\text{rms}^2=\DELT \mean{\utLpsq}\left(\der{y\tU}y\right)^2+\frac12\DELT \mean{\utLpsq}\dder{y^2\tu_\text{rms}^2}y+
\DELT \mean{\theta^2}\left(\tw_\text{rms}^2-\tu_\text{rms}^2\right)+\dots, 
\end{equation}
\begin{equation}\label{eqn:dvrms}
\Delta v_\text{rms}^2= \frac{\DELT\mean{\utLpsq}}2\dder{y^2\tv_\text{rms}^2}y+\dots, \\
\end{equation}
\begin{equation}
\label{eqn:dwrms}
\Delta w_\text{rms}^2=\frac{\DELT \mean{\utLpsq}}2\dder{y^2\tw_\text{rms}^2}y+\DELT \mean{\theta^2}\left(\tU^2+\tu_\text{rms}^2-\tw_\text{rms}^2\right)+\dots,
\end{equation}
\begin{equation}\label{eqn:duv}
\DELT \mean{u'v'}=\frac{\DELT\mean{\utLpsq}}2\dder{y^2\tilde \tau_R}y-\frac12\mean{\theta^2}\tilde \tau_R +\dots. 
\end{equation}
If the terms we identified to be small are dropped, this becomes
\begin{equation}\label{eqn:apprmoments}
\Delta  u_\text{rms}^2\approx\DELT \mean{\utLpsq}\left(\der{y\tU}y\right)^2\!\!, \ 
\Delta w_\text{rms}^2\approx \DELT \mean{\theta^2}\tU^2, \ 
\Delta v_\text{rms}^2\approx \DELT \mean{u'v'}\approx 0.
\end{equation}

\revision{Comparing the expressions (\ref{eqn:LS}) for the large-scale components with the expressions for the velocity mean squares (\ref{eqn:urmsF}) and (\ref{eqn:wrmsF}) and their increments  
(\ref{eqn:durms})  and (\ref{eqn:dwrms}) shows that the non-zero terms in (\ref{eqn:apprmoments}) are the contributions of the large-scale motion. Thus, (\ref{eqn:apprmoments}) is in agreement with the well-known observations, see for example \citep{baidya2012}, to which (\ref{eqn:apprmoments}) only adds the form of functional dependence between these increments and the shape of the mean velocity profile.}

\subsection{Comparisons with numerical simulations and experiment}

The prediction of the QSQH theory will be compared with the results of the direct numerical simulations by \cite{hoyas06} and \cite{LeeMoserJFM2015} and of the experiments by \cite{baidya2012}. These results are presented in wall units. All the comparisons will be done for the case of the small amplitude of the fluctuation of large-scale motions:  $\mean{\utLpsq}\ll1$ and $\mean{\theta^2}\ll1$. The relationships that will be verified by comparisons keep only the main term of the expansions in powers of  $\mean{\utLpsq}$ and $\mean{\theta^2}$. As the relationships (\ref{eqn:wallunits}) between the units of the QSQH theory and the wall units show, these units coincide in the main term of the expansion. Moreover, for the values of $\Rey_\tau$ considered the difference between these units is too small to be noticeable in plots. To reflect this, the wall units notation will be used in the plots, but the QSQH notation will be kept in the text.  

Some of the comparisons will require the knowledge of the universal mean profile  and of the second moments of the universal velocity. From (\ref{eqn:urmsT}-\ref{eqn:uv})
it follows that in the main term of the expansion they coincide with the mean and the second moments of the actual velocity. These parameters do not vary significantly within the range of $\Rey_\tau$ for which the comparisons will be made, but for consistency we will always use the DNS data for $\Rey_\tau=5200$ as describing the universal velocity, that is we will assume that $(\tU,  \tu_\text{rms}^2, \tv_\text{rms}^2,\tw_\text{rms}^2,\tilde \tau_R)$=$\left.(U,u_\text{rms}^2, v_\text{rms}^2,w_\text{rms}^2, \mean{u'v'})\right|_{\Rey_\tau=5200}$. Each comparison involves two flows at different $\Rey_\tau$, which will be specified. 

\revision{
\subsubsection{Logarithmic derivatives of mean square of wall parallel velocity components}
}

Multiplying by $y$ the logarithmic derivative of  (\ref{eqn:apprmoments}) with respect to $y$ gives
\begin{equation}\label{eqn:apprmomentsU}
D_u(y)=y\der{}y{\ln \Delta  u_\text{rms}^2}\approx 2y\der{}y{\ln\der{y\tU}y}, \ 
\end{equation}
\begin{equation}\label{eqn:apprmomentsW}
D_w(y)=y\der{}y\ln\Delta w_\text{rms}^2\approx 2y\der{}y\ln\tU.
\end{equation}

\begin{figure}
\setlength{\unitlength}{1 ex}
\hrule
	\centering\begin{picture}(122,30)

\put(3,2){\includegraphics[height=25 ex, clip, trim = 0 0 0 0]{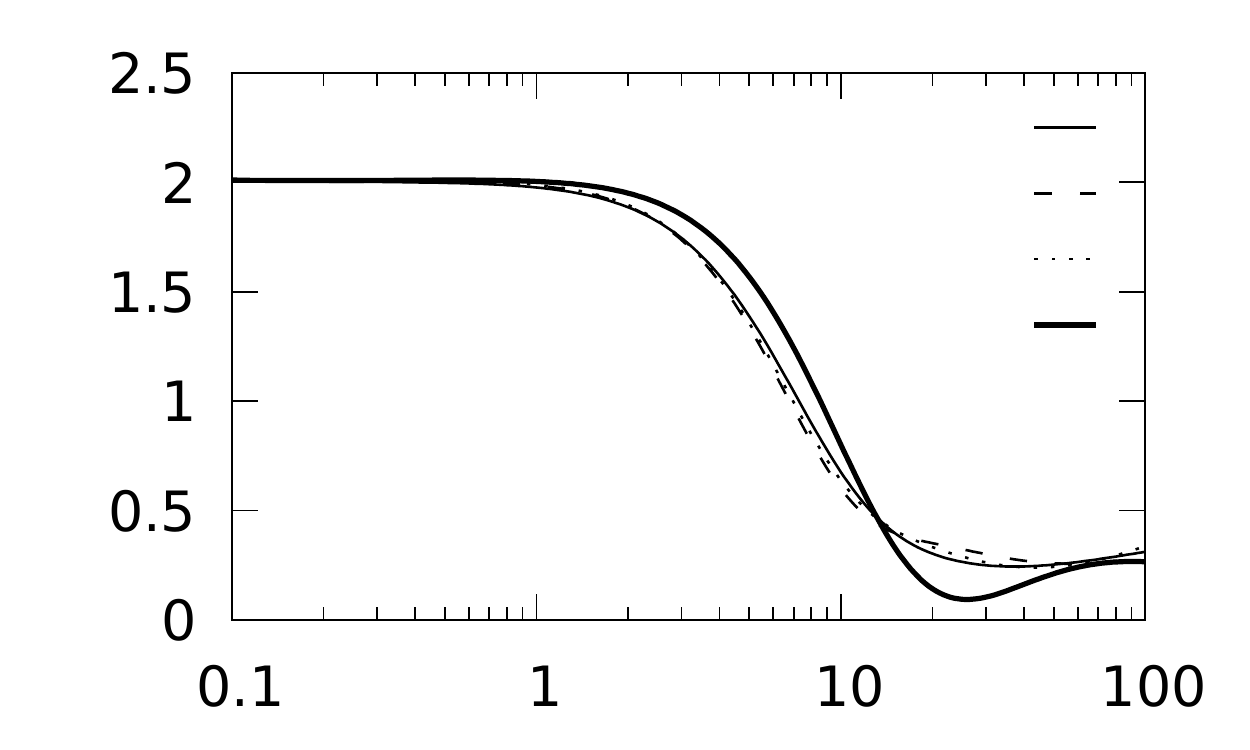}}
\put(12,22.3){\small (a)}
\put (2,15){$D_u$}
\put(24.3,22.3){\tiny \bf $\Rey_\tau$: 2000-5200}
\put(28.5,20.1){\tiny \bf 1000-2000}
\put(28.8,18.0){\tiny \bf \ 950-2000}
\put(31.5,15.7){\tiny \bf QSQH }

\put(47,2){\includegraphics[height=25 ex, clip, trim = 0 0 0 0]{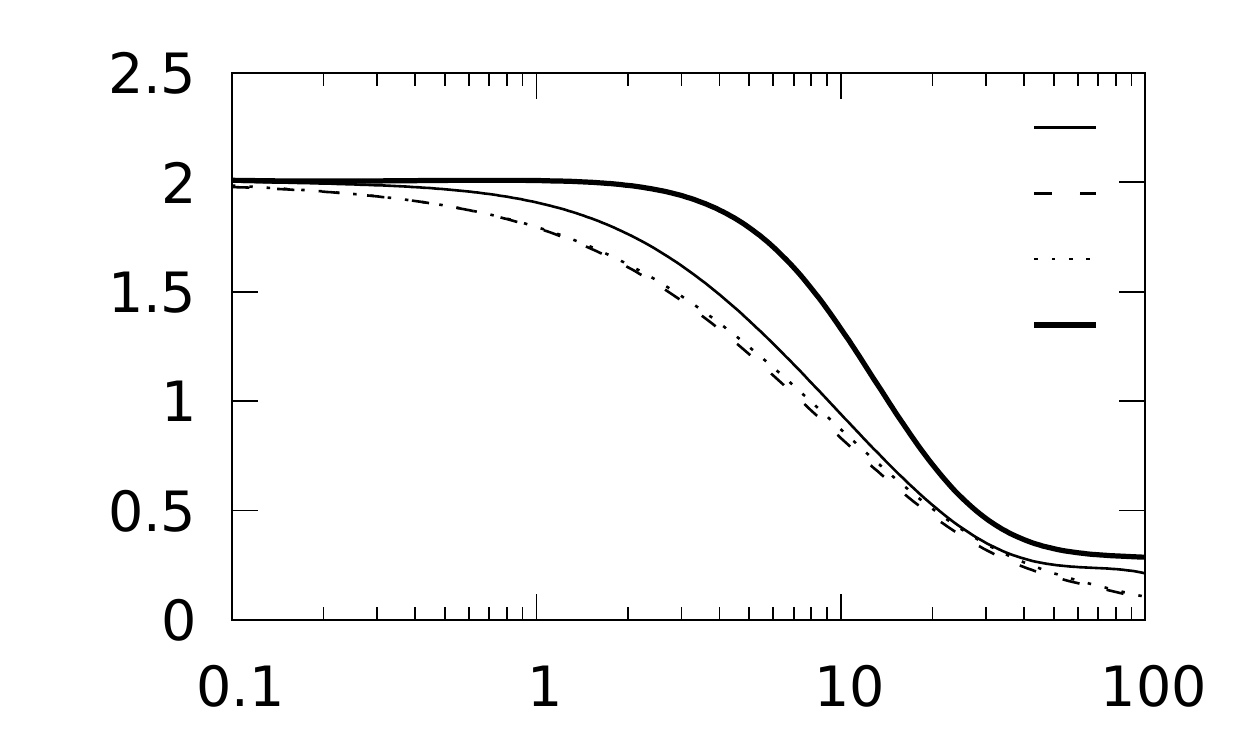}}

\put(56,22.3){\small (b)}
\put (47,15){$D_w$}
\put(68.3,22.3){\tiny \bf $\Rey_\tau$: 2000-5200}
\put(72.5,20.1){\tiny \bf 1000-2000}
\put(72.8,18.0){\tiny \bf \ 950-2000}
\put(75.5,15.6){\tiny \bf QSQH }

\put(25,0.5){$y^+$}
\put(70,0.5){$y^+$}

\end{picture}
	\caption{ Comparison of the logarithmic derivatives of the increments of $u_\text{rms}^2$ (a) and $w_\text{rms}^2$ (b) as $\Rey_\tau$ varies. Thin  curves are DNS results. The QSQH predictions are (\ref{eqn:apprmomentsU}--\ref{eqn:apprmomentsW}). The increments correspond to the changes in $\Rey_\tau$ as shown.
 \label{fig:LogsOfDeltas}}
	\end{figure}
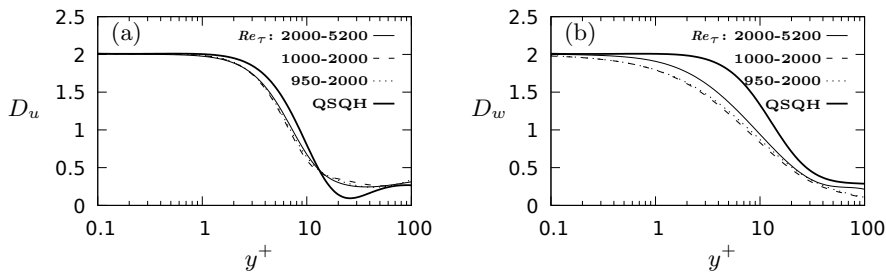

\revision{
Figure~\ref{fig:LogsOfDeltas} shows the comparison for  $D_u$ and $D_w$. Figure~\ref{fig:LogsOfDeltas}(a) 
is effectively the same as figure~12 in \citep{Zhang16}, with the curve corresponding to the increment in $\Rey_\tau$ from 950 to 2000 added, and the quality of the DNS data fit improved, in particular in the area of small values of $y^+$. It is worth to mention that since the value of the $y$-multiplied logarithmic derivative of any function proportional to ${y}^2$ as $y\to0$ is 2, the very good agreement for very small values of $y$ should not be overrated.
In figure~\ref{fig:LogsOfDeltas}b close to $y^+=100$, 
the deviation of the curves for the differences between the flows with  $\Rey_\tau$ of 950 and 2000 and of 1000 and 2000  is due to the imperfection of the functional fit to the DNS data, aggravated by differentiation. 
Figure~\ref{fig:LogsOfDeltas}(b) shows that the agreement between DNS and the QSQH theory noticeably improves as $\Rey_\tau$ increases. This is expected, but figure~\ref{fig:LogsOfDeltas}(b) is the first definite confirmation that the accuracy of the three-dimensional QSQH hypothesis 
 improves with increasing $\Rey_\tau$. The improvement can also be noticed in figure~\ref{fig:LogsOfDeltas}(a), although in that figure the agreement is noticeably better, while the improvement is smaller.
}

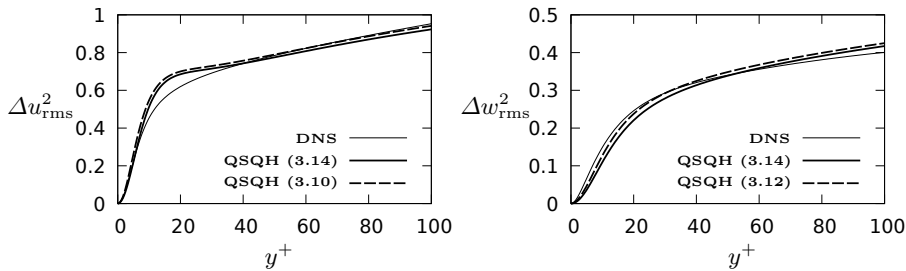
\begin{figure}
\setlength{\unitlength}{1 ex}
	\centering\begin{picture}(122,30)

\put(3  ,2){\includegraphics[height=25 ex, clip, trim = 0 0 0 0]{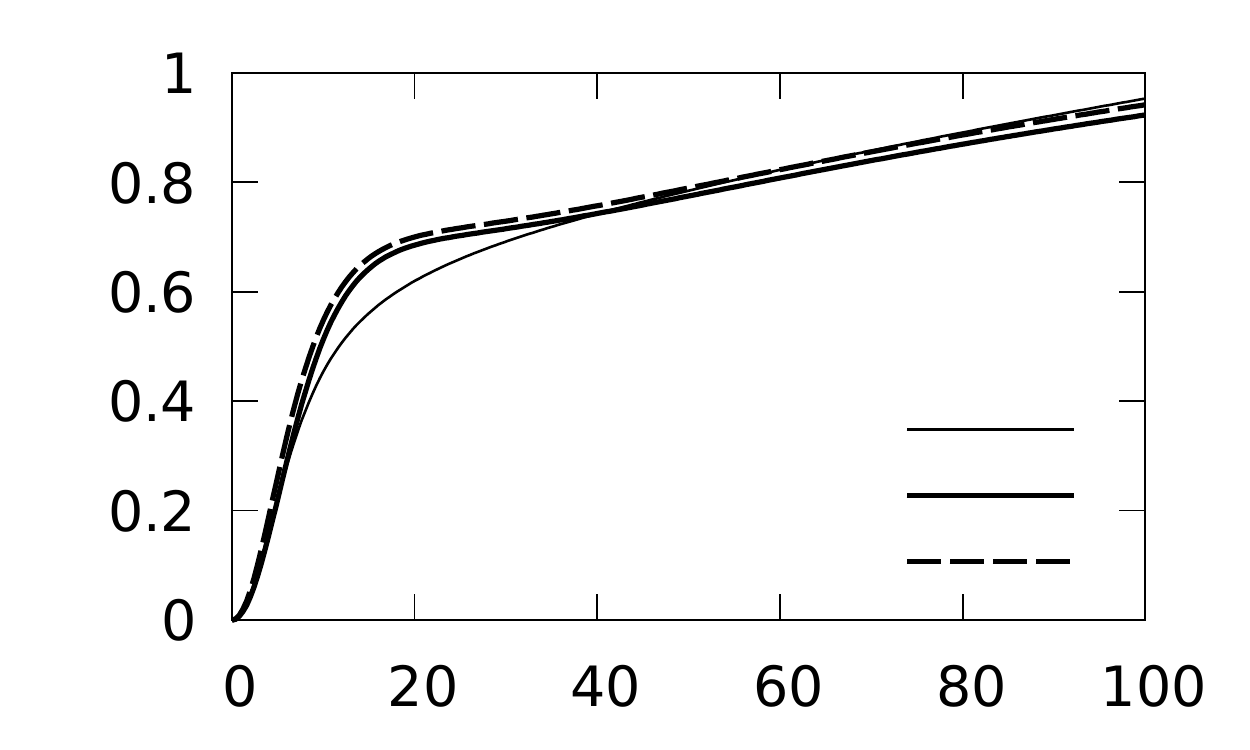}}
\put(47,2){\includegraphics[height=25 ex, clip, trim = 0 0 0 0]{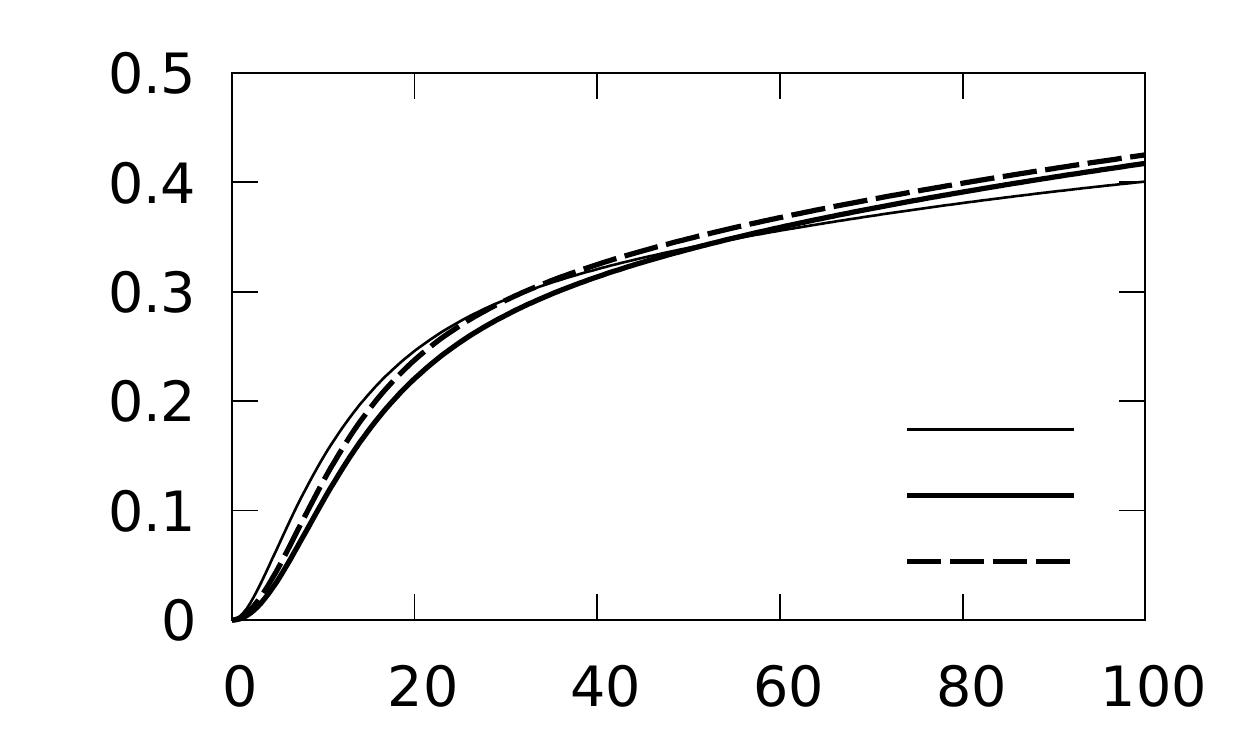}}
\put (0,15){$\Delta u_\text{rms}^2$}
\put (44,15){$\Delta w_\text{rms}^2$}


\put (28,12.2){{\tiny \bf DNS}}
\put (21,10.1){{\tiny \bf QSQH  (\ref{eqn:apprmoments})}}
\put (21,8){{\tiny \bf QSQH  (\ref{eqn:durms})}}

\put (72,12.2){{\tiny \bf DNS}}
\put (65,10.1){{\tiny \bf QSQH  (\ref{eqn:apprmoments})}}
\put (65,8){{\tiny \bf QSQH  (\ref{eqn:dwrms})}}

\put(25,0.5){$y^+$}
\put(70,0.5){$y^+$}
%

%
\end{picture}
	\caption{Comparison of the increments of 
$u_\text{rms}^2$ and $w_\text{rms}^2$ as $\Rey_\tau$ changes from 2000 to 5200. The thin  curves are DNS results. The solid curves are QSQH predictions with simplified formulae (\ref{eqn:apprmoments})  and the dashed curves are QSQH predictions with the full formulae
(\ref{eqn:durms})
and (\ref{eqn:dwrms}).
 The fitting parameters  $\DELT\mean{\utLpsq}$ and $\DELT\mean{\utLpsq}$ are given in table~\ref{table}.
 \label{fig:deltaUWs}}
	\end{figure}

\revision{

 Comparison using logarithmic derivatives cannot be extended to the full formulae (\ref{eqn:dU}--\ref{eqn:duv}). Fortunately, it is still possible to compare if not the magnitude but at least the shape of the curves of the increments of the second moments themselves rather than their derivatives. To do this, we fit $\Delta  u_\text{rms}^2(y)$ and $\Delta w_\text{rms}^2(y)$ with the expressions  (\ref{eqn:apprmomentsU}--\ref{eqn:apprmomentsW}), using $\DELT\mean{\utLpsq}$ and  $\DELT\mean{\theta^2}$ as the fitting parameters.} 
\revision{Fitting was done with the gnuplot command `fit', which implements the 
nonlinear least-squares Marquardt-Levenberg algorithm. The obtained values of $\DELT\mean{\utLpsq}$ and  $\DELT\mean{\theta^2}$ are listed in table~\ref{table}. With these values, $\Delta  u_\text{rms}^2(y)$ and $\Delta w_\text{rms}^2(y)$ can be found not only from (\ref{eqn:apprmoments}) but also from (\ref{eqn:durms},\ref{eqn:dwrms}). Both results are shown in figure~\ref{fig:deltaUWs}.
}
%
 Overall, the deviation of the QSQH predictions for $\Delta  u_\text{rms}^2$ and $\Delta w_\text{rms}^2$ is not large and is of the same order of magnitude as reported in most cases in \citep{Zhang16}.  This further confirms that near the wall the variation of $  u_\text{rms}^2$ and $  w_\text{rms}^2$ with $\Rey$ is dominated by the QSQH mechanism.
\revision{
 It is interesting to note that while figures~\ref{fig:LogsOfDeltas} and~\ref{fig:deltaUWs} are comparisons for the same QSQH formulae, the visual impression  is that in figure~\ref{fig:LogsOfDeltas} the predictions involving  $\Delta  u_\text{rms}^2$ are more accurate than the prediction involving $\Delta w_\text{rms}^2$ while vice versa in figure~\ref{fig:deltaUWs}. 
}

\begin{table} 
    \begin{center}
      \begin{tabular}{cccc}
       ${\Rey_\tau}_1$            &     {${\Rey_\tau}_2$} & {$\DELT\mean{\utLpsq}$}  & {$\DELT\mean{\theta^2}$}\\[3pt]
      \hphantom{0}950                         &      2000                  & 0.002739                               & 0.001480\\
       1000 & 2000    & 0.002286 & 0.001273\\
       2000 &  5200 & 0.002611 & 0.001544
      \end{tabular}
    \end{center}
  \caption{Increments of  $\mean{\utLpsq}$ and $\mean{\theta^2}$ as $\Rey_\tau$ grows from ${\Rey_\tau}_1$ to ${\Rey_\tau}_2$, fitted to (\ref{eqn:apprmoments})\label{table}. Four digits are accurate only for the specific fits to the specific set of data as described in the text. }
  \end{table}

 The small difference between the QSQH predictions for $\Delta  u_\text{rms}^2$ and $\Delta w_\text{rms}^2$ obtained using (\ref{eqn:apprmoments}) and (\ref{eqn:durms},\ref{eqn:dwrms}) confirms that the terms dropped in (\ref{eqn:apprmoments}) are small.


The use of fitting has a drawback. The QSQH theory relies on the assumption that there exists a large-scale filter satisfying the properties  P1-P5 (\ref{eqn:Li}-\ref{eqn:Lv}). The good agreement in figure~\ref{fig:deltaUWs}  does not guarantee with full certainty, however, that the values of $\DELT \mean{\utLpsq}$ and $\DELT\mean{\theta^2}$ obtained with such filter and the values obtained by fitting are close. The comparisons in
 figure~\ref{fig:LogsOfDeltas} are free from this complication, because, with $\tU\approx U$,   (\ref{eqn:apprmomentsU}) and (\ref{eqn:apprmomentsW}) establish a flow property  that is independent of the definition or even existence of the large-scale filter.  Another informative comparison of such nature can be done by introducing functions $R_u(y)$ and $R_w(y)$ as
\begin{equation}\label{eqn:RuRw}
R_u(y)=\frac{\Delta u_\text{rms}^2}{\left(\der{y\tU}y\right)^2\left(1+C_u(y)\right)}, \quad R_w(y)=\frac{\Delta w_\text{rms}^2}{U^2\left(1+C_w(y)\right)},
\end{equation}
where
\begin{equation}\label{eqn:durmsRatioB}
C_u(y)=\left.\left(\frac12\dder{y^2\tu_\text{rms}^2}y+
\frac{  \DELT \mean{\theta^2} }{  \DELT \mean{\utLpsq}}
          \left( \tw_\text{rms}^2-\tu_\text{rms}^2\right)+\dots \right)  \middle/  \left( \der{y\tU}y \right)^2 \right., 
\end{equation}
and
\begin{equation}
\label{eqn:dwrmsRatioB}
C_w(y) = \left.\left(\frac{\DELT \mean{\utLpsq}}{2\DELT \mean{\theta^2}}\dder{y^2\tw_\text{rms}^2}y+\tu_\text{rms}^2-\tw_\text{rms}^2\right)\middle/ U^2\right.+\dots.
\end{equation}

Then (\ref{eqn:durms}) and (\ref{eqn:dwrms})  state that $R_u(y)=\DELT \mean{\utLpsq}$ and $R_w(y)=\DELT \mean{\theta^2}$. The same follows from  (\ref{eqn:apprmoments}) with the additional assumption of that $C(y)_u=C_w(y)=0$. In fact,  $C_u(y)$ and $C_w(y)$ are just the ratios of the terms neglected to the terms kept when  (\ref{eqn:apprmoments}) was obtained from (\ref{eqn:durms}) and (\ref{eqn:dwrms}). Therefore, the QSQH theory predicts that for the small amplitude of the fluctuation of the large-scale motions functions $R_u(y)$ and $R_w$(y) are independent of $y$. Moreover, with $C_u$ and $C_w$ neglected, these functions do not depend on the large-scale filter or any other elements of the QSQH theory: this is a prediction about the properties of the flow itself. 

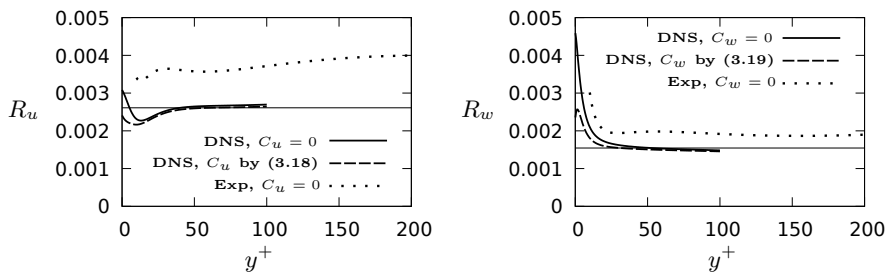
\begin{figure}
\setlength{\unitlength}{1 ex}
	\centering\begin{picture}(122,30)

\put(3  ,2){\includegraphics[height=25 ex, clip, trim = 0 0 0 0]{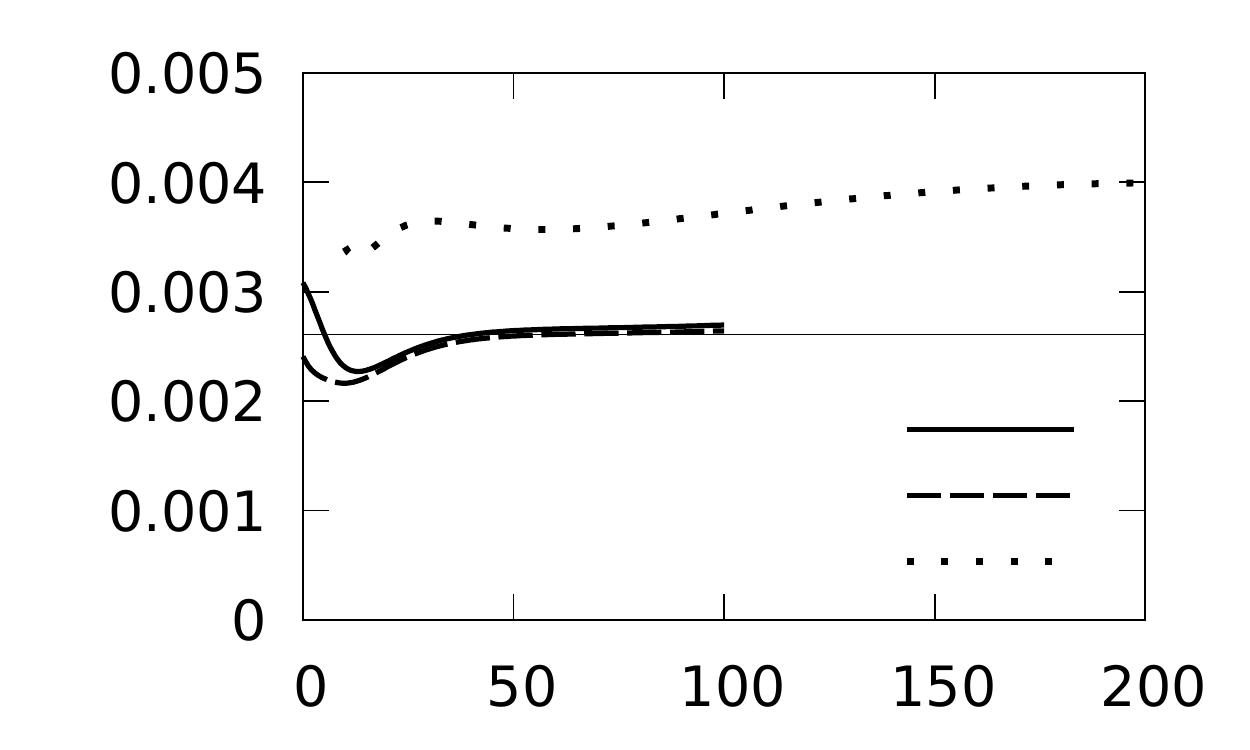}}
\put(47,2){\includegraphics[height=25 ex, clip, trim = 0 0 0 0]{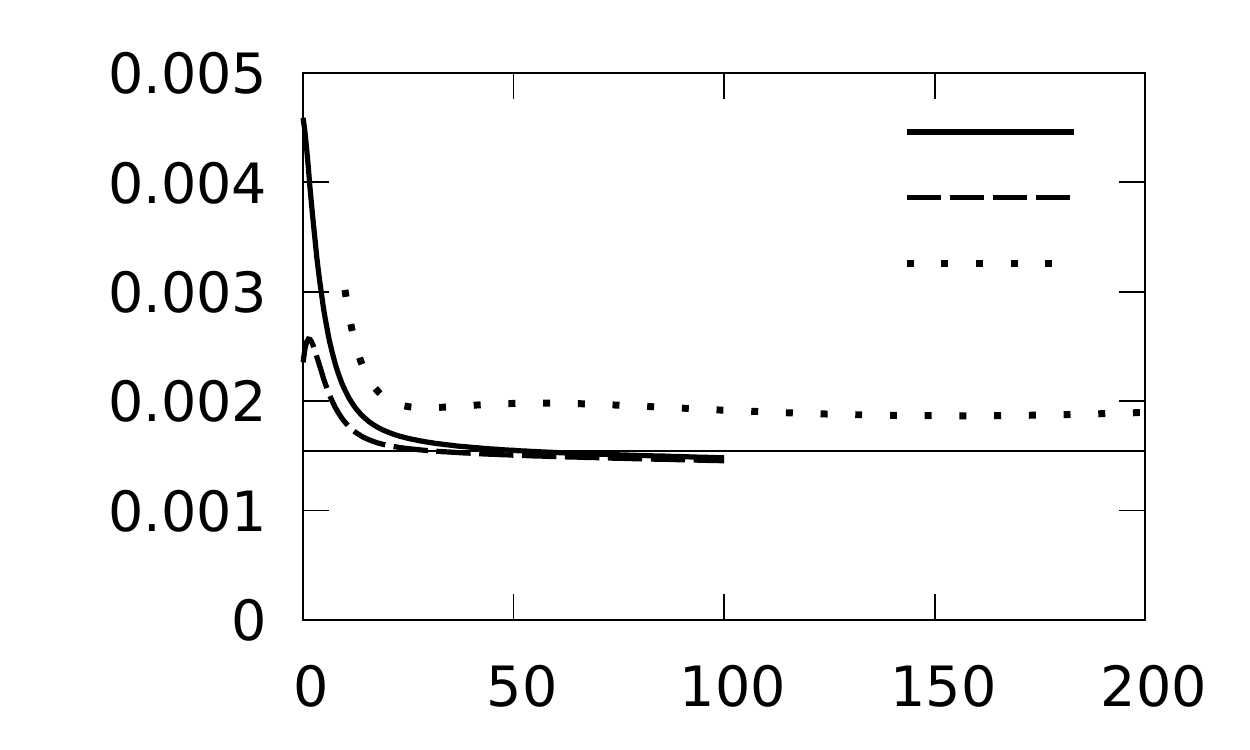}}
\put (2,15){$R_u$}
\put (46,15){$R_w$}

\put (22,8){{\tiny \bf Exp, $C_u=0$}}
\put (16,10.1){{\tiny \bf DNS, $C_u$ by (\ref{eqn:durmsRatioB})}}
\put (21,12.2){{\tiny \bf DNS, $C_u=0$}}

\put (66,18){{\tiny \bf Exp, $C_w=0$}}
\put (60,20.1){{\tiny \bf DNS, $C_w$ by (\ref{eqn:dwrmsRatioB})}}
\put (65,22.2){{\tiny \bf DNS, $C_w=0$}}

\put(25,0.5){$y^+$}
\put(70,0.5){$y^+$}
%

%
\end{picture}
	\caption{Comparison of the functions $R_u(y^+)$ and $R_w(y^+)$, (\ref{eqn:RuRw}), with DNS and experimental data. In the limit of small amplitude of the fluctuation of the large-scale motions the QSQH theory predicts that $R_u$ and $R_w$ are independent of $y^+$.  The DNS results are for channel flows with $\Rey_\tau$ changing from 2000 to 5200, and the experimental results are for a boundary layer with $\Rey_\tau$ changing from 3000 to 10500.   
 The solid and dotted curves are for $C_u=C_w=0$. The dashed curves use the full formulae (\ref{eqn:durmsRatioB}) and (\ref{eqn:dwrmsRatioB}). The horizontal straight lines are the values of $\DELT \mean{\utLpsq}$ and $\DELT \mean{\theta^2}$ from table~\ref{table}.
 \label{fig:deltaUWratios}}
	\end{figure}

 Plots of $R_u(y)$ and $R_w(y)$ are shown in figure \ref{fig:deltaUWratios}.  The DNS data are for the difference of the mean velocity squares in flows in a plane channel  with $\Rey_\tau=2000$ and $5200$  \citep{hoyas06}. The data range up to 100 wall units is determined by the range of our fits to the mean square velocities, which in turn was selected on the basis of previous findings \citep{Zhang16, agostini_leschziner_2019}  that at least for $\Rey_\tau=1000$ the QSQH predictions deviate from the the DNS results for $y^+>100$. 
The experimental data are for a zero pressure gradient boundary layer with  $\Rey_\tau=3000$ and $10500$  \citep{baidya2012}. For these higher $\Rey_\tau$ the comparison data range was
 extended to $y^+=200$. The data were obtained by digitizing  the large-scale contribution points in figure~5 in~\citep{baidya2012}, and then using the same curve-fitting procedure as described
 in the supplement for the DNS data. The large-scale contribution was used because the variation of the small-scale contribution is small, as evidenced by that figure itself, and as predicted by the QSQH theory, and because the comparison was made only for $C_u=C_w=0$, that is for the QSQH predictions for the large-scale contribution (see  the discussion following (\ref{eqn:apprmoments})).
 The curves are flat for larger values of $y^+$, which corresponds to the QSQH prediction, but, \revtwo{predictably,} have noticeable variation near the wall, \revtwo{where $C_u$ and $C_w$ are not negligible}. Figures~\ref{fig:LogsOfDeltas}-\ref{fig:deltaUWratios} are various ways of verifying the same predictions (\ref{eqn:durms}), (\ref{eqn:dwrms}), and (\ref{eqn:apprmoments}). The seeming discrepancy in the performance of the QSQH theory is due to these figures describing different measures of the error.
 Thus, near the wall figure~\ref{fig:LogsOfDeltas} effectively compares the exponent of $y^+$ of the $y^+\to 0$ asymptotics of (\ref{eqn:durms}), (\ref{eqn:dwrms}), and (\ref{eqn:apprmoments}), figure~\ref{fig:deltaUWs} compares the absolute error, and figure~\ref{fig:deltaUWratios} effectively compares the relative error. Fortunately, comparisons for the other two moments,   $v_\text{rms}^2$ and $\mean{u'v'}$,  give further insight into the physics of the question.

\begin{figure}
\setlength{\unitlength}{1 ex}
	\centering\begin{picture}(122,30)

\put (22,22.1){{\tiny \bf $\Rey_\tau=1000-2000$}}
\put (66,22.1){{\tiny \bf $\Rey_\tau=2000-5200$}}

\put (22,12.1){{\tiny \bf DNS}}
\put (22,10){{\tiny \bf QSQH  (\ref{eqn:duv})}}
\put (22,7.9){{\tiny \bf $y\Delta\der px$}}

\put (66,12.1){{\tiny \bf DNS}}
\put (66,10){{\tiny \bf QSQH  (\ref{eqn:duv})}}
\put (66,7.9){{\tiny \bf $y\Delta\der px$}}

\put(25,0.5){$y^+$}
\put(70,0.5){$y^+$}

\put(3,2){\includegraphics[height=25 ex, clip, trim = 0 0 0 0]{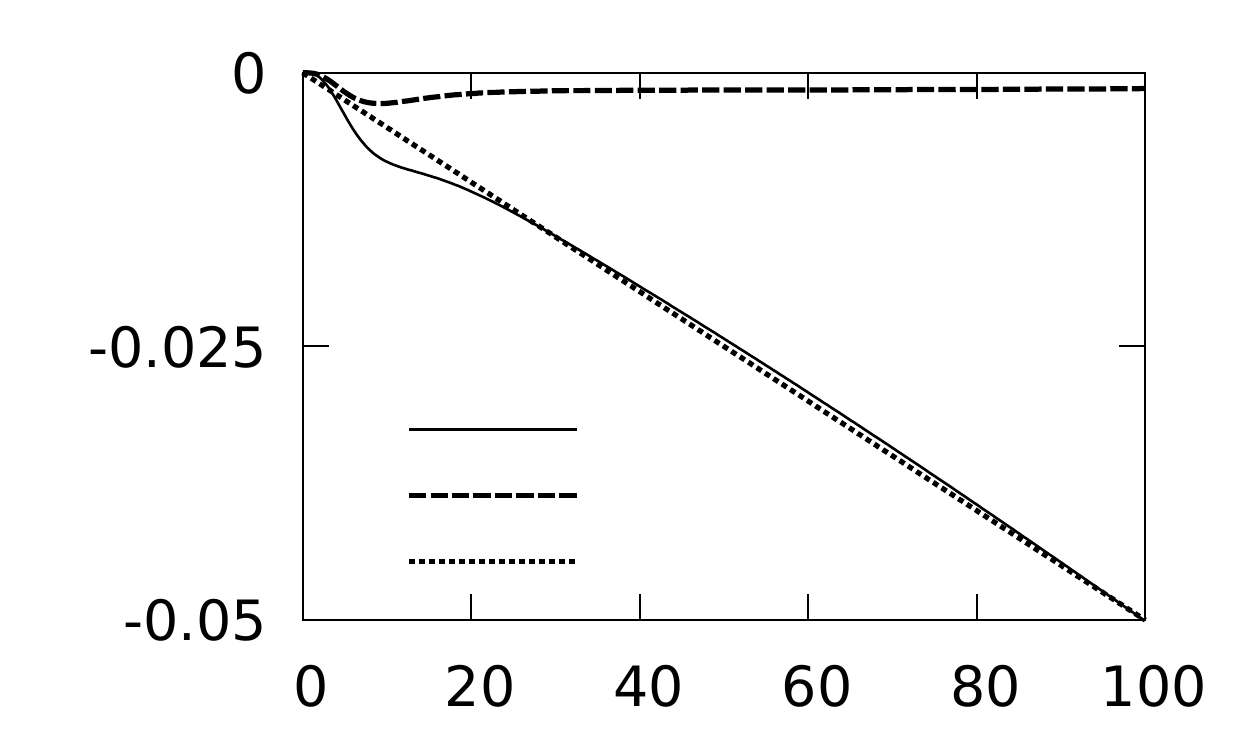}}
\put(47,2){\includegraphics[height=25 ex, clip, trim = 0 0 0 0]{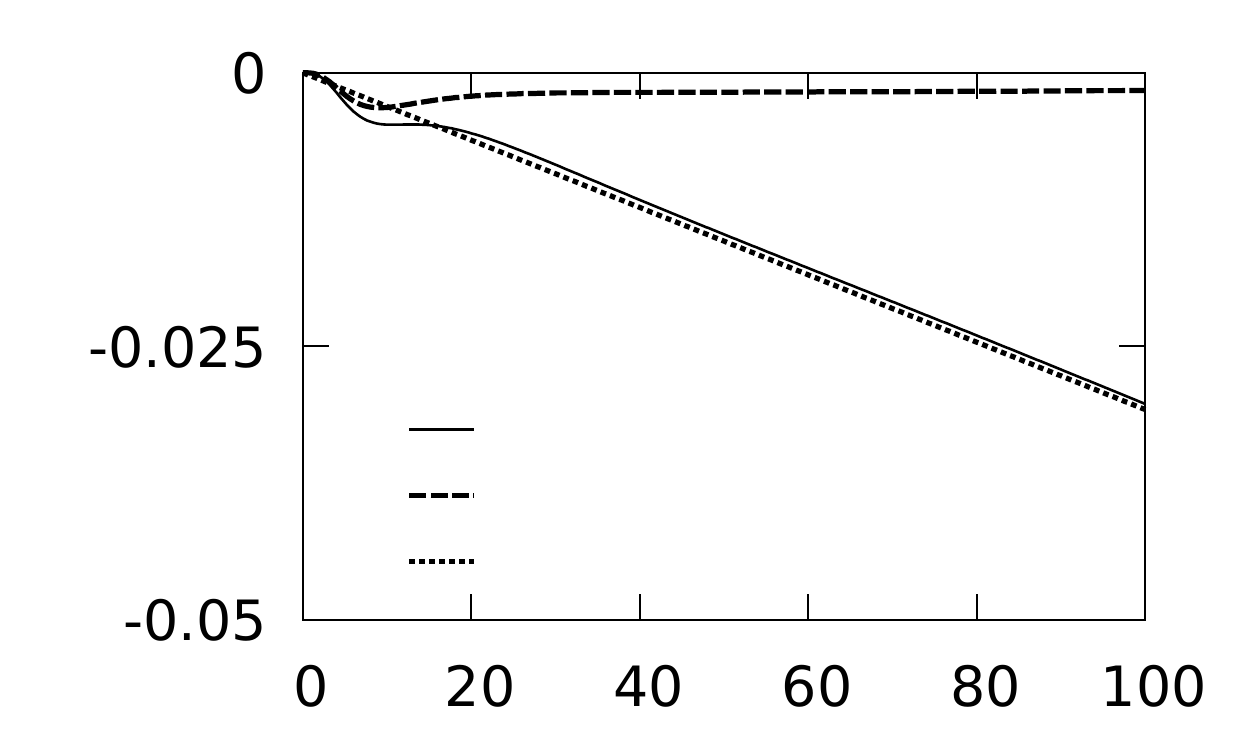}}

\put (3,19){$\DELT \mean{u'v'}$}
\put (47,19){$\DELT \mean{u'v'}$}

\end{picture}
	\caption{Comparison of the increments of $\mean{u'v'}$ as $\Rey_\tau$ changes from 1000 to 2000 (left) and from 2000 to 5200 (right). Thin  curves are DNS results. The QSQH predictions are   (\ref{eqn:duv}), shown with dashed curves.  The dotted curve is $y^+\Delta dp^+\!\!/dx^+.$  The fitting parameters  $\DELT\mean{\utLpsq}$ and $\DELT\mean{\utLpsq}$ are given in table~\ref{table}.
 \label{fig:deltaTau}}
	\end{figure}
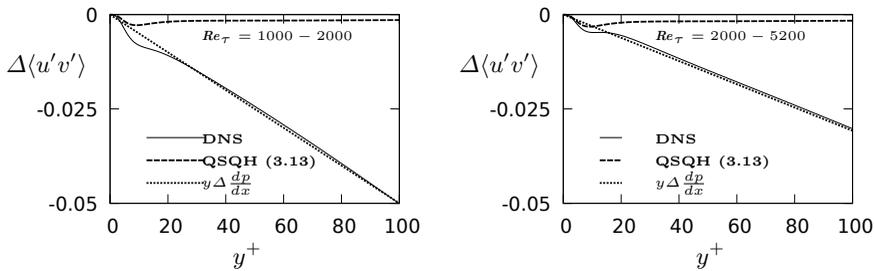

The large-scale motions contribute to $\Delta v_\text{rms}^2$ and $\DELT \mean {u'v'}$ only indirectly by modifying the small-scale structures. If this effect is neglected, as in (\ref{eqn:apprmoments}), the predicted increments of $\Delta v_\text{rms}^2$ and $\DELT \mean {u'v'}$ are zero. Accordingly, the comparisons for these quantities are made only for the full predictions.  
The QSQH predictions for $\DELT \mean {u'v'}$ deviate considerably, by an order of magnitude, from the DNS results, as shown in figure~\ref{fig:deltaTau}. The almost linear behaviour of $\DELT \mean {u'v'}$ for $y^+>20$ suggests an explanation. 
 It is well-known that the dependence of the total shear stress on the distance to the channel wall is linear and in wall units is equal to $y^+dp^+/dx^+=y^+/\Rey_\tau,$ and that further away from the wall the Reynolds stress approaches the total stress.  The corresponding change in the total stress, equal to
 $y^+\Delta dp^+dx^+=y^+(1/\Rey_{\tau_2}-1/\Rey_{\tau_1}),$ is shown with a dotted line. It closely agrees with the  DNS result everywhere except the vicinity of the wall, where both the QSQH effect and the pressure gradient effect are of the same order of magnitude. As $\Rey_\tau$ changes, so does the pressure gradient. Comparing the curves for different $\Rey_\tau$ in figure~\ref{fig:deltaTau} suggests that QSQH modulation effects can become more significant than the effect of the pressure gradient only at much higher  $\Rey_\tau$.

\begin{figure}
\setlength{\unitlength}{1 ex}
	\centering\begin{picture}(122,30)

\put (16,18){{\tiny \bf $\Rey_\tau=1000-2000$}}
\put (60,18){{\tiny \bf $\Rey_\tau=2000-5200$}}

\put (22,22.1){{\tiny \bf DNS}}
\put (22,20){{\tiny \bf QSQH  (\ref{eqn:dvrms})}}

\put (66,22.1){{\tiny \bf DNS}}
\put (66,20){{\tiny \bf QSQH  (\ref{eqn:dvrms})}}

\put(25,0.5){$y^+$}
\put(70,0.5){$y^+$}

\put(3,2){\includegraphics[height=25 ex, clip, trim = 0 0 0 0]{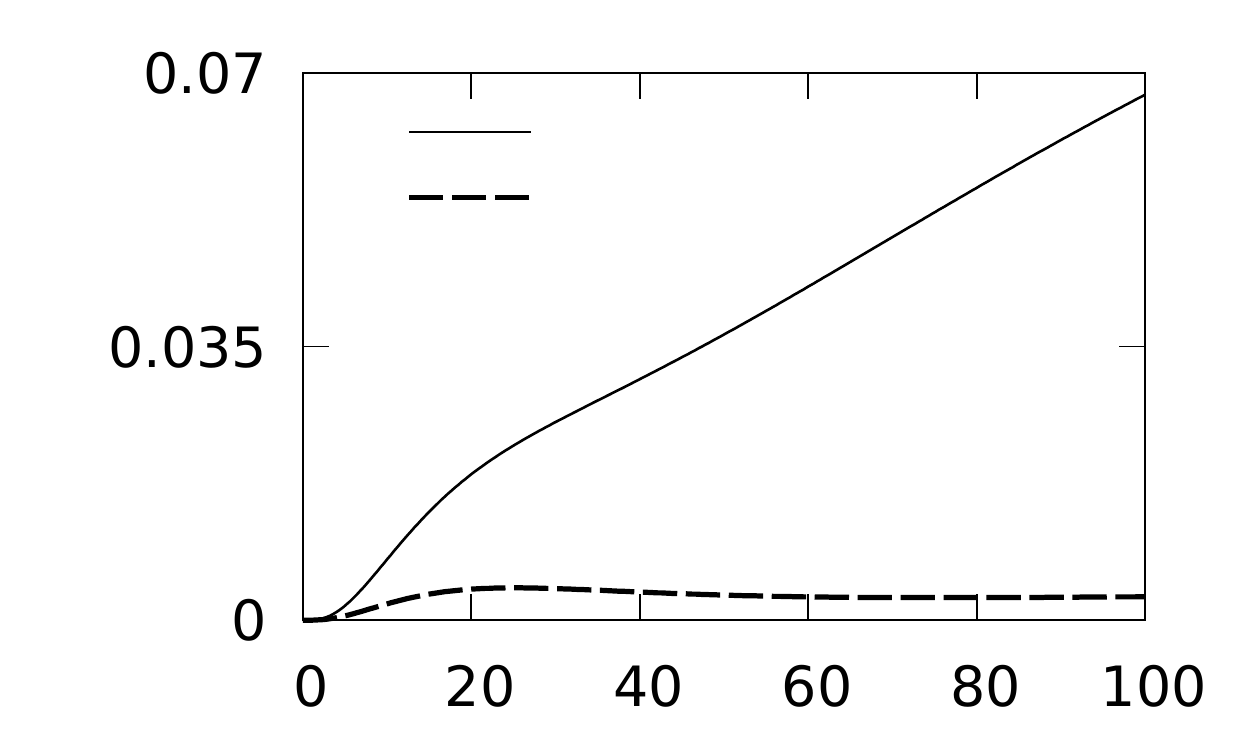}}
\put(47,2){\includegraphics[height=25 ex, clip, trim = 0 0 0 0]{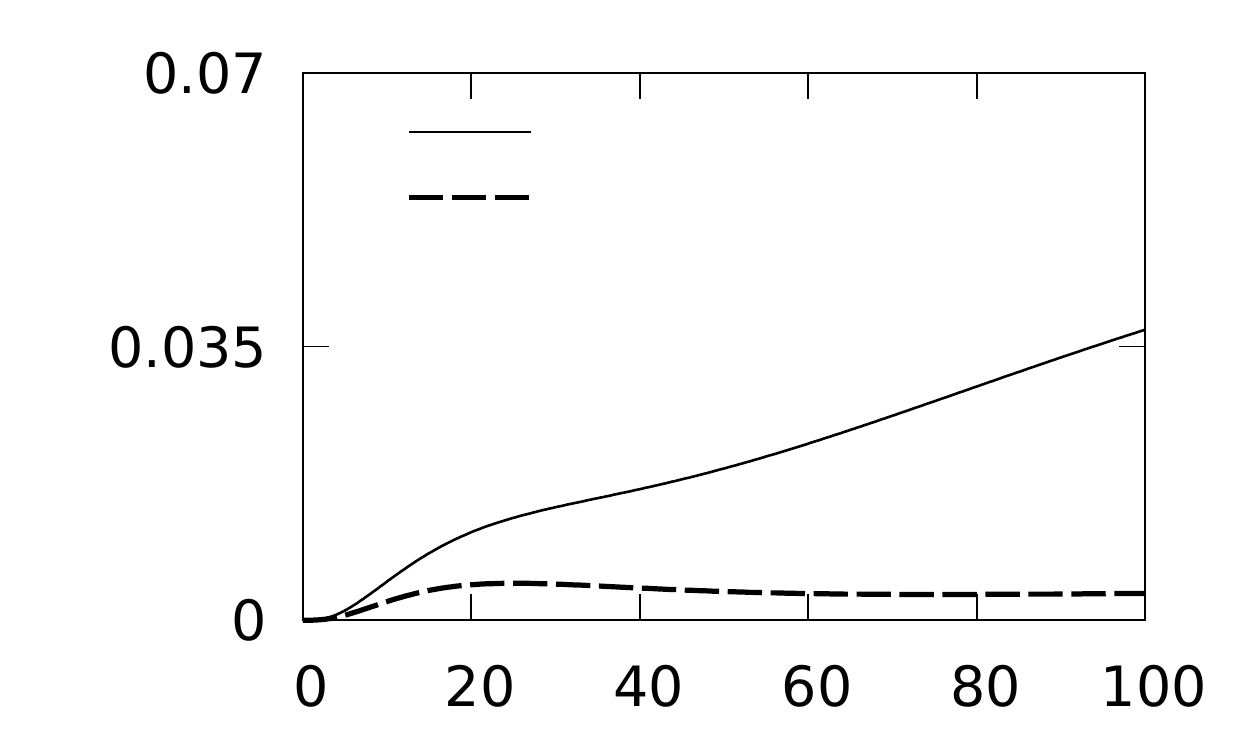}}

\put (0,14){$\Delta v_\text{rms}^2$}
\put (44,14.5){$\Delta v_\text{rms}^2$}

\end{picture}
	\caption{Comparison of the increments of $v_\text{rms}^2$  as $\Rey_\tau$ changes from 1000 to 2000 (left)
and from 2000 to 5200 (right). Thin  curves are DNS results. The QSQH predictions are (\ref{eqn:dvrms}).    The fitting parameters  $\DELT\mean{\utLpsq}$ and $\DELT\mean{\utLpsq}$ are given in table~\ref{table}.
 \label{fig:deltaVrms}}
	\end{figure}

Comparisons made in figure~\ref{fig:deltaVrms} for $\Delta v_\text{rms}^2$ also reveal a significant deviation of the QSQH predictions from the DNS results.  There is no simple formula for the quantitative description of the effect of the pressure gradient on $v_\text{rms}$, but since both  $v_\text{rms}$ and $ \mean {u'v'}$  depend on the properties of the wall-normal velocity, it is reasonable to assume that their similar behaviour has the same nature. This is further confirmed by the decrease of the deviation as $\Rey_\tau$ increases. 

The  comparisons in the section draw a clear physical picture. The near-wall turbulent flows are affected both by the large-scale structures and other effects, unrelated to large-scale structures, such as the pressure gradient in a channel flow. In the range of the $\Rey_\tau$ considered, that is primarily from $1000$ to $5000$, these mechanisms can create effects of various magnitude depending on the particular statistic in question and the distance to the wall. When the effect of the large-scale structures dominates, it can be described by the QSQH theory.

\section{Non-universality of the constants of the logarithmic law}\label{sec:loglaw}

The well-known logarithmic  law states that for large $\Rey$ there is an interval of the distances from the wall where the mean velocity profile can be described by the formula \citep{townsend1980structure,PopeS.B2000Tf}
\begin{equation}\label{eqn:loglaw}
U^+(y^+)=\frac1\kappa\ln y^+ + B.
\end{equation}
This law is intrinsically linked to the classical universality hypothesis, and it can be obtained from a simple dimensional analysis. Therefore, if the classical universality hypothesis is invalidated by the effects of large-scale motions on the near-wall flows, so should be the logarithmic law.  Note that the justification based on dimensional analysis applies in equal measure to the universality of the constants $\kappa$ and $B$ and to the functional, logarithmic, form of it.
 Therefore, it can be expected that the influence of large-scale motion on the near-wall turbulence  invalidates the functional form of the shape of the profile, thus making the question of the universality of the logarithmic law constants redundant.
 \cite{Zhang16} noticed that, remarkably, if the influence of the large-scale motions is described by the QSQH theory then it does make the constants not universal while nevertheless conserves the functional logarithmic dependence of the mean profile on the distance to the wall. Here we first will extend this observation to the case when the QSQH theory takes into account the fluctuations of the direction of the large-scale friction.   

The same arguments leading to the logarithmic behaviour of $U^+(y^+)$ within the classical universality hypothesis can be applied to $\tU(\ty)$ within the QSQH theory, giving  
\begin{equation}\label{eqn:univloglaw}
\tU(\ty)=\frac1{\tilde \kappa}\log \ty + \tilde B.
\end{equation}
Substituting (\ref{eqn:U}) and using the wall units (\ref{eqn:wallunits}) gives after simple transformations
\begin{equation}\label{eqn:newloglaw}
U^+(y^+)=\frac{\mean{\utL^+\cos\theta}}{\tilde\kappa}\log y^+ +\mean{\left(\frac1{\tilde \kappa}\log\utL^+ +\tilde B\right)\utL^+\cos\theta}.
\end{equation}
Comparing with (\ref{eqn:loglaw}) shows that
\begin{equation}\label{eqn:logconst}
\kappa=\frac{\tilde\kappa}{\mean{\utL^+\cos\theta}}\quad \text{ and }\quad B=\mean{\left(\frac1{\tilde \kappa}\log\utL^+ +\tilde B\right)\utL^+\cos\theta}.
\end{equation}
(Note the missing `+' superscript of $B$ in the formula following (10) in~\citep{Zhang16}. That formula can  be obtained from (\ref{eqn:newloglaw}) by setting $\theta=0$, that is by neglecting the fluctuations of the direction of the large-scale component of the skin friction.)

The large-scale statistics entering (\ref{eqn:logconst}) depend on the characteristics of the large-scale motions, so that the constants can be expected to be different for pipe, channel, boundary layer, or other-geometry flows, and for different $\Rey$ (as far as the logarithmic law applies at finite $\Rey$). The attempts to estimate the constants in experiments and numerical calculations do give different values for different conditions, but the observed differences are not large. This can be due to the relatively small variation of the amplitude of the fluctuations of the large-scale motions.       
For $\mean{\utLpsq}\ll 1$ and $\theta\ll1$ the Taylor expansion of (\ref{eqn:logconst}) gives
\[
\kappa=\tilde\kappa\left(1-\frac{\mean{\utLpsq}}2+\frac{\mean{\theta^2}}2+\dots\right),
\]
\[
B=\tilde B -\left(\tilde B -1\right)\frac{\mean{\utLpsq}}2-\tilde B \frac{\mean{\theta^2}}2+\dots.
\]
Within this approximation the difference between the values of the constants in two different flows can be expressed in terms of the differences in $\mean{\utLpsq}$ and  $\mean{\theta^2}$  as
\[
\Delta \kappa=\tilde\kappa\left(-\frac{\Delta\mean{\utLpsq}}2+\frac{\Delta\mean{\theta^2}}2+\dots\right),
\]
\[
\Delta B= -\left(\tilde B -1\right)\frac{\Delta\mean{\utLpsq}}2-\tilde B \frac{\Delta\mean{\theta^2}}2+\dots.
\]
Recollecting that according to (\ref{eqn:apprmoments})
\[
\frac{\Delta  u_\text{rms}^2}{\left(\der{y\tU}y\right)^2}\approx\DELT \mean{\utLpsq}\!\!, \ 
\frac{\Delta w_\text{rms}^2}{\tU^2}\approx \DELT \mean{\theta^2}
\]
gives 
\begin{equation}\label{eqn:dkappa}
\frac{\Delta \kappa}{\kappa}\approx -\frac{\Delta  u_\text{rms}^2}{2\left(\der{y\tU}y\right)^2}+\frac{\Delta w_\text{rms}^2}{2\tU^2}
\end{equation}
and
\begin{equation}\label{eqn:dB}
\frac{\Delta B}B\approx  -\left(1 -\frac1B\right)\frac{\Delta  u_\text{rms}^2}{2\left(\der{y\tU}y\right)^2}- \frac{\Delta w_\text{rms}^2}{2\tU^2}.
\end{equation}
Combining these two one can also obtain a relationship not involving the spanwise velocity:
\begin{equation}\label{eqn:dkappaB}
\frac{\Delta \kappa}{\kappa}+\frac{\Delta B}B\approx \left(\frac1B-2\right)\frac{\Delta  u_\text{rms}^2}{2\left(\der{y\tU}y\right)^2}.
\end{equation}
Now, we can estimate the approximate magnitude of the variation of the constants caused by the effect of large-scale motions on the near-wall region. In particular, from figure~\ref{fig:deltaUWratios} for the experimental data of \cite{baidya2012} follow crude estimates  ${\Delta  u_\text{rms}^2}/{\left(\der{y\tU}y\right)^2}\approx 0.04$ and  ${\Delta w_\text{rms}^2}/{2\tU^2}\approx 0.02$ for $\Rey_\tau$ changing from $3000$ to $10500$. Then (\ref{eqn:dkappa}) and (\ref{eqn:dB}) give a 1\% variation of $\kappa$ and even smaller variation of $B$. This is within the current level of uncertainty of experimental measurements \citep{marusic2013logarithmic}, but not too small to eliminate the hope that the accuracy sufficient to verify (\ref{eqn:dkappa}) experimentally will be achieved in the foreseeable future. 





\section{The effect of the fluctuations of the direction of large-scale skin friction on the conditional averages of spanwise velocity}
\label{sec:OnAgostiniLeschziner}

For the channel flow at $\Rey_\tau=1000$, \citet{agostini_leschziner_2019}  plotted (in different notation)
 the conditional
 average
$\hat w^2(\hat y,{\utL}_x)=\left.\mean{(w\left(y{\utL}_x)/{\utL}_x\right)^2}\right|_{{\utL}_x}$
 versus  $\hat y=y{\utL}_x$ for several values of ${\utL}_x=\utL\cos\theta$. 
Unlike other parameters, the curves of $\hat w(\hat y,{\utL}_x)$ for different values of  ${\utL}_x$ deviated from each other to a certain degree even quite close to the wall. 
If the fluctuations of the direction of the large-scale component of skin friction were negligible (that is if $\theta=0$), $\hat w^2$ would coincide with $\tw^2,$ and according the QSQH assumption, the curves should be independent of $\utL$. 
There are three possible reasons for the deviation:  neglecting the fluctuations of the large-scale component of the wall friction,  the imperfection of the large-scale filter used, and the approximate nature of the QSQH hypothesis.
 We will consider first how taking into account that $\theta\ne0$ affects this issue. \citet{agostini_leschziner_2019} mentioned that 
 using for scaling the large-scale friction velocity $\utL$ instead of its $x$-component does not make much difference.
%
%
%
Conditionally averaging the square of (\ref{eqn:QSQHonePointw}) for fixed $\utL$ and using the filter properties and symmetry gives 
\begin{equation}\label{eqn:Dtildew}
 \left.\mean{\frac{w^2}{\utL^2}}\right|_{\utL}=
\left(\tilde U^2(\ty)+\tilde u^2_\text{rms}(\ty)-\tilde w^2_\text{rms}(\ty)\right)
\left.\mean{\sin^2\theta}\right|_{\utL}+\tilde w^2_\text{rms}(\ty).
\end{equation}
%
%
%
Therefore,  $\left.\mean{{w^2}/{\utL^2}}\right|_{\utL}$ varies with $\utL$ in proportion to the variation of $\left.\mean{\sin^2\theta}\right|_{\utL}$.
Since 
$\tU^2$ is greater than the other terms, (\ref{eqn:Dtildew}) implies that the increment of $\left.\mean{{w^2}/{\utL^2}}\right|_{\utL}$ as $\utL$ varies has the same sign for all values of $\ty.$
 However, the shape of the curves of $\left.\mean{{w^2}/{\utL^2}}\right|_{\utL}$ in  figure~4(f) in~\citep{agostini_leschziner_2019} shows that in fact  this increment changes the sign at approximately $\ty=70.$
 Hence, if the QSQH theory were expected to be valid for $\ty\approx y^+ > 70$, accounting for the fluctuation of the direction of the large-scale wall friction would not explain the discrepancy.
 For $\ty < 70$ taking into account the fluctuations of the large-scale friction direction might improve the collapse of the curves of $\left.\mean{{w^2}/{\utL^2}}\right|_{\utL}$ provided that $\left.\mean{\sin^2\theta}\right|_{\utL}$suitably  increases with $\utL.$

The deviation from the QSQH theory might also be caused by the properties of the bi-dimensional EMD filter used in \citep{agostini_leschziner_2019}. Indeed, the EMD filter does not satisfy properties P1, P3, and P4 of the QSQH filter. This, for example, invalidates the application of (\ref{eqn:Dtildew}) to the results of \citep{agostini_leschziner_2019}. On the other hand, the EMD filter can be expected to satisfy to a reasonable degree the property P5, since both the EMD and the conventional Fourier cut-off filter were  found to be effective in separating the scales ~\citep{Dogan2018revisiting}.  A deeper analysis is complicated. The large-scale motions and the small-scale motions obtained using the bi-dimensional EMD filter are not guaranteed to satisfy continuity. Hence, there is an exchange of mass, momentum and energy between the large and small scales thus defined. 
\revision{
This does not necessarily invalidates the application of this filter in QSQH theory. In fact, the large-scale velocity field given by (\ref{eqn:LS}) is also not guaranteed, and is in fact unlikely, to satisfy continuity.  However, the deviations from continuity of large-scale motions of the form of  (\ref{eqn:LS})  and of the large-scale motions defined by the EMD filter might be different.
}
Also, the small-scale motions defined using the EMD filter are not guaranteed to have zero average and appear to actually have a nonzero average (see  figure 1(b) in \citep{agostini_leschziner_2019}). As a result, even an intuitive reasoning about the results obtained using this filter is hard. 
We have to postpone further judgment until a thorough analysis of the above features of the bi-dimensional EMD filter becomes available.




\section{Discussion}

At the start of this discussion it is appropriate to make a note on terminology. The QSQH theory describes the effect of the large-scale motions on the near-wall region of a turbulent flow. In the context of two input signals resulting in one output signal the term superposition refers to the case when the output signal is the sum of  one signal and the other signal multiplied by a constant. The term amplitude modulation refers to the case when the output signal is the product of one signal and an amplitude, which is a function of the other signal. 
If the signals are functions of time, the term frequency modulation refers to the case when the output signal  is the first signal with the time argument being replaced by the time multiplied by a function of the other signal. 
When the signal is a function of a spacial coordinate, one can also refer to scale modulation that is similar to the frequency modulation but with a spatial coordinate instead of time. Distinguishing between these forms of signal interaction is useful, as modulation and superposition often have different physical mechanisms. The first quantitative empirical relations describing the influence of large-scale structures on the near-wall region of a turbulent flow had the form of a combination of amplitude modulation and superposition \citep{mathis09a}.  
  In contrast, the QSQH theory involves only amplitude, frequency, and scale modulation, but not superposition.  \cite{ChernyshenkoEtAl:2012:arxiv} pointed out that the empirical relation of   \cite{mathis09a} follows from the QSQH theory in the limit of small amplitude of the fluctuations of large-scale motion, and that the superposition term in the empirical relation is just a linearised representation of both amplitude and wall-normal scale modulation effects. In what follows we will describe all these effects with one generic term modulation.   

Modulation of near-wall turbulence by large-scale motions is only one of many effects inherent in near-wall turbulent flows.  
\revtwo{As $\Rey$ increases, the modulation effect becomes progressively stronger~\citep{mathis09a}. This naturally follows from the assumption that the intensity of the outer motions has outer or at least mixed scaling~\citep{Marusic10}, from which it follows that as $\Rey$ tends to infinity, the strength of the modulation effect also tends to infinity. While the assumption about the limiting behaviour can be questioned~\citep{chen_sreenivasan_2021}, the trend itself is well established for a wide range of the Reynolds numbers (see for example figure 2 in~\citep{hultmark2012turbulent}). It is also well known that  different effects have different scaling. Thus, the near-wall cycle scales in wall units (see for example the recent results of~\cite{Baidya2017} or infer this from the relative success of  the majority of semi-empirical turbulence models based on the assumption of pure wall scaling of near-wall turbulence). Yet other effects, not only the pressure gradient in a channel flow considered specifically in the present study, but other effects such as, for example, the effect of the nonzero inertial terms of the mean velocity in a zero-pressure-gradient boundary layer, became progressively weaker (in wall units) as $\Rey$ increases.  

The results obtained in the present study demonstrate that the modulation effect can dominate other effects or be comparable or negligible as compared to them  depending on the quantity of interest being considered. That in turbulent flows the same physical effect can have different importance for different quantities of interest has already been suggested in the early work of \cite{townsend_1961}, who envisaged different scaling of `active' and `inactive' motions he introduced. The results presented here exemplify this in the context of the QSQH theory: the QSQH large-scale motions are inactive.   


The present study shows that certain effects that can be expected to become negligible as compared to modulation as $\Rey$ tends to infinity, remain important in the range of $\Rey_\tau$ considered, that is at least up to about 5000, and are likely to remain important for much higher $\Rey_\tau$. In general, this is not surprising. For example, \cite{luchini2017universality} demonstrated that the effect of the pressure gradient on the shape of the mean velocity profile remains so significant at $\Rey_\tau=1000$ that taking it into account   gives a noticeable improvement in detecting the logarithmic layer.  
}

The QSQH theory describes only a part of the modulation effects. First, it relies on the assumptions similar to the assumptions of the classical  universality theory. These assumptions become strictly valid only as $\Rey$ tends to infinity.
 In addition, the QSQH theory relies on the separation of scales. Since turbulent flows are characterised by a continuos spectrum, it is not obvious that a complete separation of scales can be achieved even in the limit of infinite $\Rey$. Separation of scales is clearly not perfect in the range of $\Rey$ of  the majority of current  experiments and numerical simulations.
The situations when the effects described by the QSQH theory are likely to dominate the other effects can be detected from within the theory itself, as in section~\ref{sec:momentssensitivity}, or from comparisons, as could be done from observing the behaviour of $R_u(y)$ and $R_w(y)$ in the experimental data in figure~\ref{fig:deltaUWratios}.

Within the QSQH theory Taylor expansions in the magnitude of the fluctuation of the large-scale motions combined with algebraic manipulations can give results  independent of the definition of the  large-scale filter. In the present study this is exemplified by $D_u$ (\ref{eqn:apprmomentsU}),  $D_w$ (\ref{eqn:apprmomentsW}), $R_u$ and $R_w$ (\ref{eqn:RuRw}) being constant when $C_u\approx C_w\approx 0$,  and the formulae for the constants of the logarithmic law (\ref{eqn:dkappa}-\ref{eqn:dkappaB}). This allows comparisons to be made even when the large-scale components were not measured or calculated. This also opens a possibility to a new form of analysis of the existing data.

It is important to note that since $\tu$ is the same in all flows,  in principle the QSQH relationships between the increments of various statistics should apply to any pair of high-Reynolds number turbulent flows, not necessarily different by the values of $\Rey_\tau$.
 For example, they should apply if one flow is a flow in a pipe while another is the flow in a flat-plate boundary layer.  Comparisons of such nature are yet to be made. 
\revtwo{ It is instructive to note the difference between the universality we refer to here and the classical universality. For example, in the work by \cite{luchini2017universality} we just mentioned, the difference between the shape of the mean profiles in a pipe and plane channel was explained by the difference in the effect of the pressure gradient. In wall units this effect tends to zero, recovering the classical universality as $\Rey_\tau\to\infty$, which is one of the conclusions in~\citep{luchini2017universality}. In contrast, the large-scale motions in a pipe and in a plane channel can be expected to remain different at arbitrary large Reynolds numbers, and the QSQH theory will then express the difference in the near-wall flow in terms of the properties of the large-scale motions, including the difference in the shape of the mean profile and the constants of the logarithmic law. However, as it follows from the results of Section~\ref{sec:loglaw}, this difference can be expected to be noticeable only at higher Reynolds numbers.  
}

 The QSQH mechanisms of the modulation of near-wall turbulence by large-scale motion have a rigorous mathematical description. They can be explored by means of rigorous mathematics independently of numerical simulations and experiments. This rigorous consideration does not, however, apply to the other, non-QSQH, mechanisms of modulation, which have therefore  to be explored by other means. Since the QSQH mechanisms are often not negligible and can even be dominant, they can obscure other mechanisms. This obstacle can be overcome by using the variables of the QSQH theory, namely $\tilde t$, $\tilde \bx$, $\tilde \bu$ as defined by (\ref{eqn:QSQHxyz}-\ref{eqn:tilde_variables}). If the results obtained numerically or experimentally are converted to these `tilde' variables, the QSQH effects will be automatically eliminated from the consideration. Using these variables required the knowledge of $\utL$ and $\theta$, which it turn implies that the large-scale filter is defined. Fortunately, in such an application the choice of the specific filter might be not crucial, as far as the filter is reasonable. The reasons for this are discussed in \citep{Chernyshenko_FDR2019}.  An example of this is presented in figures~4,5, and 9 in~\citep{agostini_leschziner_2019}. In these figures the collapse of the data in `tilde'  variables is much better than the collapse of data presented in standard wall units, even though the EMD filter used does not satisfy several of the properties of the ideal QSQH filter and even though the fluctuations of the direction of the large-scale friction were not taken into account.      

Using the tilde variables for the analysis of  experimental data has an additional difficulty of the need to obtain both the magnitude and the direction of the large-scale friction. This can be done either by measuring the friction itself or by measuring the velocity very close to the wall, which represents a serious difficulty at high Reynolds numbers. In this respect it should be pointed out that the large-scale component needs to be filtered from the total field.
  The success of this operation depends on the signal-to-noise ratio, with noise in this instance being the small-scale fluctuations. The magnitude of the small-scale fluctuation reduces as the distance from the wall increases beyond the position of the near-wall maximum, while the magnitude of large-scale fluctuations increases. Hence, the measurement of the large-scale component is better to be done at a certain not too-small distance from the wall. Crucially, provided that the region of the measurement is still within the range of validity of the QSQH approximation, measurements of the large-scale velocity components can be used for calculating $\utL$ and $\theta$ by solving the set of equations (\ref{eqn:LS}) in which $\tU\approx U$, and $\utL$ and $\theta$ are considered as the unknowns. Judging by figure~\ref{fig:deltaUWratios}, for the current range of $\Rey_\tau$ characteristic of experiments the reasonable choice of the distance from the wall, where such measurement are to be made, is about 100 wall units. A suggestion of a probe for measuring the large-scale components in such conditions is made in \citep{Chernyshenko_FDR2019}.

\section{Conclusions}

The extension of the QSQH theory proposed in the present paper takes into account the fluctuations of the direction of the large-scale component of the wall friction. It describes the effect of large-scale motions on all three components of the velocity of near-wall turbulent flows. \revtwo{This distinguishes the present work from \citep{ChernyshenkoEtAl:2012:arxiv} and \citep{Zhang16}.} The theory was applied to several questions related to the near-wall turbulent flows and gave the following specific results. 

It is found that the effect of the fluctuation of the direction of the large-scale component of wall friction is significant. \revtwo{In other words, the effect of modulation of the near-wall turbulence by the spanwise component of the large-scale motions is as significant, both quantitatively and conceptually, as the modulation by the streamwise component.}

The extended theory explains the large sensitivity of the magnitudes $u_\text{rms}$ and $w_\text{rms}$ of the fluctuations of longitudinal and spanwise velocities to the variation in the Reynolds number in comparison with the sensitivity of the mean velocity profile, the Reynolds stress, and the magnitude of the fluctuations of the wall-normal velocity. 
It is shown that the variation of $u_\text{rms}$ with $\Rey$ is largely caused by the variation of the amplitude and wall-normal-scale modulation by the outer, large-scale, $\Rey$-dependent motions, while the variation of $w_\text{rms}$ is largely caused by the fluctuations of the direction of the large-scale, $\Rey$-dependent, component of the wall friction. 

Explicit relationships between the difference in the second moments of velocity in any two high-Reynolds-number near-wall flows were derived, and comparisons were made for flows with different Reynolds number. The comparisons gave a satisfactory agreement for $u_\text{rms}$ and $w_\text{rms}$ in the range of the distances from the wall where the modulation by large-scale motions dominates other effects.


\revtwo{The theory shows} that for a channel flow the variation of the Reynolds stress as $\Rey_\tau$ varies in the range below 5000 and probably well above is dominated by the variation of the pressure gradient and not by the variation in the large-scale structures. A similar conclusion was found to be likely also for  $v_\text{rms}$.

Relationship between the differences of the constants of the logarithmic law, the shape of the mean velocity profile, and the differences of the second moment caused by the differences in  large-scale motions were derived. Using experimental data for $\Rey_\tau=3000$ and $\Rey_\tau=10500$, the difference in $\kappa$ due to the difference in the large-scale motions in these two cases was estimated to be about 1\%, which is somewhat below the current error margins of experimental measurements. 

\medskip

Declaration of Interests. The author reports no conflict of interest.

\appendix
\section{Detailed derivations}

Derivation of (\ref{eqn:U}):
\begin{multline*}
U(y)\bydef=
\mean{u(y)}\byQSQHu= 
\mean{\utL\tu(y\utL)\cos\theta} +\mean{\utL \tw(y\utL)\sin\theta}\\
\byP{1,2,4}=
\mean{\LSF\utL\tu(y\utL)\cos\theta} +\mean{\LSF\utL\tw(y\utL)\sin\theta}
\byP5=
 \mean{\utL \tilde U(y\utL)\cos\theta}.
\end{multline*}

Derivation of (\ref{eqn:LS}):
Using that  $\tilde W(\ty)=\mean{\tw(\ty)}\stackrel{(\ref{eqn:ZeroMeans})}=0$ gives
\begin{multline*}
u_L(y)\stackrel{\text{def}}=\LSF u(y)\stackrel{(\ref{eqn:QSQHonePointu})}=\LSF \left(\utL\tu(y\utL)\cos\theta + \tw(y\utL)\utL \sin\theta\right)\\
\stackrel{P1,5}= \utL \tilde U(y\utL)\cos\theta.
\end{multline*}
  Similar derivations give also $v_L$ and $w_L$.

Derivation of (\ref{eqn:LScorr}):

Note that $\mean{u_S}=0$ since $U\stackrel{(\ref{eqn:U},\ref{eqn:LS})}=\mean{u_L }$. Also note that $\mean{\tu'(\ty)}_{\ty=\text{const}}=\mean{\tw(\ty)}_{\ty=\text{const}}\stackrel{(\ref{eqn:ZeroMeans})}=0$. Hence,
\[
\mean{u'_Lu'_S}=\mean{u_Lu_S}=\mean{
	\utL^2\tU(y\utL)\cos\theta\left(
		\tu'(y\utL)\cos\theta+\tw(y\utL)\sin\theta
                                              \right)}
\byP5=0.
\]
Similarly,
\[
\mean{w'_Lw'_S}=\mean{-\utL^2\tU(y\utL)\sin\theta\left(-\tu'(y\utL)\sin\theta+\tw(y\utL)\cos\theta\right)}=0.
\]
and since $v_L\stackrel{(\ref{eqn:LS})}=0$, and $\mean{\tv(\ty)}_{\ty=\text{const}}\stackrel{(\ref{eqn:ZeroMeans})}=0$,
\[
\mean{u'_L v}=\mean{\utL^2\tU(y\utL)\cos\theta\tv(y\utL}\byP5=0
\]

Derivation of (\ref{eqn:urmsF}):
\begin{multline*}
u_\text{rms}^2(y)
\bydef=\mean{(u-U)^2}=\mean{u^2}-U^2\byQSQHu=
\mean{\left(\utL(\tilde u(y\utL)\cos\theta +\tilde w(y\utL) \sin\theta)\right)^2}-U^2\\
\byP{2,4}=\mean{\LSF\left(\utL^2(\tilde u^2(y\utL)\cos^2\theta +2\tilde u(y\utL) \tilde w(y\utL) \cos\theta\sin\theta+\tilde w^2(y\utL) \sin^2\theta)\right)}-U^2\\
\bysym=\mean{\LSF\left(\utL^2(\tilde u^2(y\utL)\cos^2\theta +\tilde w^2(y\utL) \sin^2\theta)\right)}-U^2\\
\bydef=\mean{\LSF\left(\utL^2((\tU(y\utL)+\tu'(y\utL))^2\cos^2\theta +\tilde w^2(y\utL) \sin^2\theta)\right)}-U^2\\
\byP{5}=\mean{\left(\utL^2((\tU^2(y\utL)+\tu_\text{rms}^2(y\utL))\cos^2\theta +\tw_\text{rms}^2(y\utL) \sin^2\theta)\right)}-U^2=\\
\mean{\left(\utL\tU(y\utL)\cos\theta\right)^2}-U^2+\mean{\utL^2\left(\tu_\text{rms}^2(y\utL)\cos^2\theta+
\tw_\text{rms}^2(y\utL)\sin^2\theta\right)}
\end{multline*}

Derivation of (\ref{eqn:vrmsF}):
Since $\mean{v(y)}=0$,
\begin{multline*}
v_\text{rms}^2(y)
\bydef=\mean{v(y)^2} \byQSQHv=\mean{\utL^2\tilde v^2(y\utL)}
\byP{2,4}=\mean{\LSF\utL^2\tilde v^2(y\utL)}\\
\byP{5,\text{def}}=\mean{\utL^2\tv_\text{rms}^2(y\utL)}
\end{multline*}

Derivation of (\ref{eqn:wrmsF}):
\begin{multline*}
w_\text{rms}^2(y)
=\mean{(w^2(y)}\byQSQHw=
\mean{\left(\utL(\tilde w(y\utL)\cos\theta - \tilde w(y\utL) \sin\theta)\right)^2}\\
\byP{2,4}=\mean{\LSF\left(\utL^2(\tilde w^2(y\utL)\cos^2\theta -2\tilde w(y\utL) \tilde u(y\utL) \cos\theta\sin\theta+\tilde u^2(y\utL) \sin^2\theta)\right)}\\
\bysym=\mean{\LSF\left(\utL^2(\tilde w^2(y\utL)\cos^2\theta +\tilde u^2(y\utL) \sin^2\theta)\right)}\\
\byP{5}=\mean{\utL^2\left(\tw_\text{rms}^2(y\utL)\cos^2\theta +\left(\tU^2(y\utL)+\tu^2_\text{rms}(y\utL)\right) \sin^2\theta)\right)}=\\
\mean{\left(\utL\tU(y\utL)\sin\theta\right)^2}+\mean{\utL^2\left(\tw_\text{rms}^2(y\utL)\cos^2\theta+
\tu_\text{rms}^2(y\utL)\sin^2\theta\right)}
\end{multline*}

Derivation of (\ref{eqn:uvF}):
\begin{multline*}
\mean{u'(y)v'(y)}=\mean{u(y)v'(y)}
\stackrel{(\ref{eqn:QSQHonePointu},\ref{eqn:QSQHonePointv})}=
\mean{\utL^2(\tilde u(y\utL)\cos\theta +\tilde w(y\utL) \sin\theta)\tilde v(y\utL)}\\
\byP{2,4}=\mean{\LSF\left(\utL^2(\tilde u(y\utL)\cos\theta +\tilde w(y\utL) \sin\theta)\tilde v(y\utL))\right)}
\byP{5,\text{sym}}=\mean{\utL^2\tilde \tau_R(y\utL)\cos\theta}
\end{multline*}

Derivation of (\ref{eqn:Dtildew}):
\begin{multline*}
\cmean{\frac{w^2}{\utL^2}}=\left(\tilde U^2(y\utL)+\tilde u^2_\text{rms}(y\utL)\right)\cmean{\sin^2\theta}+\tilde w^2_\text{rms}(y\utL)\mean{\cos^2\theta}
\\
=\left(\tilde U^2(y\utL)+\tilde u^2_\text{rms}(y\utL)-\tilde w^2_\text{rms}(y\utL)\right)\cmean{\sin^2\theta}+\tilde w^2_\text{rms}(y\utL)
\end{multline*}

\bibliographystyle{jfm}
\bibliography{QSQH_sc}

\end{document}